\documentclass[pra,twocolumn,superscriptaddress,longbibliography]{revtex4-2}

\usepackage{amsthm, booktabs, color, mathtools, amsmath, amssymb, setspace, bbm, tensor, mathtools, placeins, graphicx, wrapfig, float, verbatim, xcolor, physics}
\usepackage[normalem]{ulem}
\usepackage{stmaryrd} %
\usepackage{units}
\usepackage{amssymb}
\usepackage{algpseudocode}

\newcommand{\prlsection}[1]{{\em {#1}.---~}}

\newcommand{\dsl}[0]{\llbracket}
\newcommand{\dsr}[0]{\rrbracket}

\newcommand{\SMLong}{Supplemental Material}

\newcommand{\SM}{SM}

\newcommand{\subfigimg}[3][,]{%
	\setbox1=\hbox{\includegraphics[#1]{#3}}%
	\leavevmode\rlap{\usebox1}%
	\rlap{\hspace*{2pt}\raisebox{\dimexpr\ht1-0.5\baselineskip}{{\bfseries \large\textsf{#2}}}}%
	\phantom{\usebox1}%
}

\newcommand{\cnotgate}{\mathsf{CNOT}}

\newcommand{\cnot}[2]{\mathsf{CNOT}\quad {#1} \quad {#2}}
\newcommand{\initX}[1]{\mathsf{InitX} \quad {#1}}
\newcommand{\initZ}[1]{\mathsf{InitZ} \quad {#1}}
\newcommand{\measX}[1]{\mathsf{MeasX} \quad {#1}}
\newcommand{\measZ}[1]{\mathsf{MeasZ} \quad {#1}}
\newcommand{\idle}[1]{\mathsf{Idle} \quad {#1}}

\newcommand{\CQT}{Centre for Quantum Technologies, National University of Singapore, 3 Science Drive 2, Singapore 117543.\looseness=-1}

\newcommand{\IHPC}{A*STAR Quantum Innovation Centre (Q.InC), Institute of High Performance Computing (IHPC), Agency for Science, Technology and Research (A*STAR), 1 Fusionopolis Way, \#16-16 Connexis, Singapore, 138632, Republic of Singapore.\looseness=-1}
\newcommand{\sutd}{Science, Mathematics and Technology Cluster, Singapore University of Technology and Design, 8 Somapah Road, Singapore 487372, Singapore}
\newcommand{\CQuERE}{Centre for Quantum Engineering, Research and Education, TCG CREST, Sector V, Salt Lake, Kolkata 700091, India.\looseness=-1}

\definecolor{THc}{rgb}{0.9,0.3,0.2}

\renewcommand{\eqref}[1]{Eq.~(\ref{#1})} %

\graphicspath{{./00_figures/}}

\def\app#1#2{%
  \mathrel{%
    \setbox0=\hbox{$#1\sim$}%
    \setbox2=\hbox{%
      \rlap{\hbox{$#1\propto$}}%
      \lower1.1\ht0\box0%
    }%
    \raise0.25\ht2\box2%
  }%
}

\ifx\proof\undefined
\newenvironment{proof}[1][\protect\proofname]{\par
	\normalfont\topsep6\p@\@plus6\p@\relax
	\trivlist
	\itemindent\parindent
	\item[\hskip\labelsep\scshape #1]\ignorespaces
}{%
	\endtrivlist\@endpefalse
}
\providecommand{\proofname}{Proof}
\fi

\newtheorem{definition}{Definition}
\newtheorem{proposition}{Proposition}

\newcommand{\idg}[1]{{\bfseries #1)}}

\providecommand{\factname}{Fact}
\providecommand{\theoremname}{Theorem}
\providecommand{\claimname}{Claim}
\providecommand{\lemmaname}{Lemma}

\setcitestyle{sort&compress,numbers}
\usepackage[unicode=true, breaklinks=false, pdfborder={0 0 1}, backref=false, colorlinks=true, linkcolor=blue, citecolor=blue]{hyperref}

\newcommand{\revA}[1]{{#1}}

\begin{document}

\title{Multivariate Bicycle Codes}
\author{Lukas Voss}
\email{lukas\_voss@icloud.com}
\affiliation{Institute of Theoretical Physics \& IQST, Ulm University, Albert-Einstein-Allee 11, D-89069 Ulm, Germany \looseness=-1}
\affiliation{\CQT}
\affiliation{Yale-NUS College, Singapore}

\author{Sim Jian Xian}
\email{simjianxian@u.nus.edu} 
\affiliation{\CQT}
\affiliation{\IHPC}

\author{Tobias Haug}
\email{tobias.haug@u.nus.edu}
\affiliation{Quantum Research Center, Technology Innovation Institute, Abu Dhabi, UAE}

\author{Kishor Bharti}
\email{kishor.bharti1@gmail.com}
\affiliation{\IHPC}
\affiliation{\CQuERE}
\affiliation{\sutd}

\begin{abstract}
Quantum error correction suppresses noise in quantum systems to allow for high-precision computations. 
In this work, we introduce Multivariate Bicycle (MB) Quantum Low-Density Parity-Check (QLDPC) codes, via an extension of the framework developed by Bravyi~\textit{et al.} [Nature, 627, 778-782 (2024)] and particularly focus on Trivariate Bicycle (TB) codes. 
Unlike the weight-6 codes proposed in their study, we offer concrete examples of weight-5 TB-QLDPC codes which promise to be more amenable to near-term experimental setups. We show that TB-QLDPC codes up to weight-6 have a bi-planar structure and often posses a two-dimensional toric layout. Under circuit level noise, we find substantially better encoding rates than comparable surface codes while offering similar error suppression capabilities.
For example, we can encode $4$ logical qubits with distance $5$ into $60$ physical qubits using weight-5 check measurements of circuit depth 7, while a surface code with these parameters requires $200$ physical qubits. 
The high encoding rate and compact layout make our codes highly suitable candidates for near-term hardware implementations, paving the way for a realizable quantum error correction protocol.  
\end{abstract}

\maketitle

 \let\oldaddcontentsline\addcontentsline%
\renewcommand{\addcontentsline}[3]{}%

\prlsection{Introduction}
Quantum computing promises to solve problems intractable for classical computers~\cite{shor1999polynomial,arora2009computational,nielsen2001quantum}. Despite advancements in improving hardware quality, noise remains the most significant challenge for realizing practical implementations to ultimately achieve both reliable and scalable quantum computations. As a consequence, quantum error correction (QEC) is required to reduce noise of the physical hardware~\cite{Mackay_1997, Resch_2021}.
Among many QEC codes~\cite{Shor_1995, Steane_1996, Kitaev_2003}, the surface code has emerged which currently leads the forefront in experimental quantum computing due to its relatively high threshold error rates~\cite{Bravyi_1998, Kitaev_2003, Fowler_2009, Fowler_2012, Zhao_2022, Asfaw_2023} and suitability for implementation with current hardware devices~\cite{Asfaw_2023}. Despite its advantages, the surface code requires high qubit overhead leading to an asymptotically zero encoding rate while the distance scales as the square-root of the number of physical qubits~\cite{Gottesman_2010}. 

Recently, Quantum Low-Density Parity-Check (QLDPC) codes have attracted attention for their potential to require fewer physical qubits than surface codes~\cite{Breuckmann_2021}. LDPC codes are well-established in classical error correction~\cite{Mackay_2003, Shokrollahi_2004, Moon_2020} for their efficiency and decoding performance. 
In a recent breakthrough~\cite{Bravyi_2024}, Bravyi \textit{et al.} presented a set of QLDPC codes that are significantly more qubit-efficient than the surface code of the same number of logical qubits. 
These codes, called Bivariate Bicycle (BB) QLDPC codes, have origins in the work by Kovalev and Pryadko~\cite{kovalev2013quantum}. 
Ref.~\cite{Bravyi_2024} introduced codes with weight-$6$ stabilizer checks which can be encoded into a bi-planar toric layout. This layout allows their code to accommodate  four local stabilizer checks and two long-range checks.
\revA{However, QLDPC codes present significant challenges for current experimental capabilities due to the required long-range connections. Considering the example of superconducting qubits, current experimental capabilities struggle to provide extensive long-range connections across the entire circuit board, although individual long-range connections have had considerable progress in their gate fidelities recently~\cite{song2024realizationhighfidelityperfectentangler}. Therefore, an experimental proof of concept for the viability of QLDPC codes should be a minimalistic progression from fully local codes such as the surface code, e.g. by requiring only one long-range connection per qubit. %
}

Here, we introduce Multivariate Bicycle (MB) QLDPC codes to provide several novel low-weight codes with better encoding rates than comparable surface codes. We define MB-QLDPC codes in a general manner, and then focus on the concrete case of Trivariate Bicycle (TB) codes, which have three variable in the generating polynomials. 
\revA{As a highlight, we find a weight-$5$ $\dsl30, 4, 5\dsr$ TB code with a single long-range check which encodes $4$ logical qubits using only $60$ physical qubits in a bi-planar toric architecture, while a comparable surface code would require $200$ qubits. We design a depth-7 syndrome measurement circuit which preserves the distance even under circuit-level noise, in contrast to conventional depth-10 circuits. Under circuit level noise, our code has comparable noise suppression to surface codes and previously proposed weight-6 BB codes~\cite{Bravyi_2024,Berthusen_2024}. We also find for our codes several fault-tolerant logical circuits that can be implemented with transversal operations and SWAPs via automorphism groups~\cite{sayginel2024fault}.}
We also make progress on the properties of TB codes by showing that TB codes up to weight-$6$ possess a bi-planar structure.
Furthermore, we provide a weight-independent criterion to check whether a TB code has a toric layout structure.  
We also show that all stabilizer checks are translationally invariant when the MB-QLDPC codes have a toric layout.
Finally, we show that weight-$4$ TB codes can possess a tangled toric layout, where the Tanner graph corresponds to a deformed torus with some mismatched edges.

\prlsection{Framework}\label{sec::framework}
We give an introduction to quantum error correction  in Appendix~\ref{app:background} and now proceed to introduce MB-QLDPC codes and its group algebra with $r$ variables $\mathbb{F}_{2}\left[G_{r}\right]$, where $G_{r}=\nicefrac{\mathbb{Z}}{l_{1}}\times\nicefrac{\mathbb{Z}}{l_{2}}\times\cdots\times\nicefrac{\mathbb{Z}}{l_{r}}$.
In the special case of $r=2$, we recover the group algebra of the
BB code. 
\begin{figure}[htbp]
    \centering
    \includegraphics[width=0.9\linewidth]{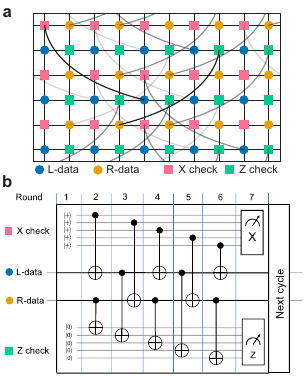}
    \caption{\idg{a} Our weight-5 $\dsl30, 4, 5\dsr$ TB-QLDPC code can be arranged in a bi-planar grid with periodic boundary conditions (PBC) using $30$ data qubits (which belong either to the L or R sublattice) and $30$ qubits for parity check measurements. Each parity check involves four neighbouring data qubits, and one non-local interaction described by a translational invariant lattice vector $(x,y)$, where $x$ is the horizontal and $y$ the vertical axis of the grid. For X-checks and L sublattice, the interaction vector is  $(-4,3)$, while for Z-checks and R sublattice we have $(4,3)$. For better visibility, we only plot representatives of the non-local interactions.
    \revA{\idg{b} Measurement syndrome circuit for our weight-5 $\dsl30, 4, 5\dsr$ TB-QLDPC with only depth 7 and circuit level distance $d_\text{circ}\leq5$, which our numerics suggest is tight.}  } 
    \label{fig:sketch}
\end{figure}
The group algebra $\mathbb{F}_{2}\left[G_{r}\right]$ is isomorphic to the multivariate polynomial quotient ring $R_{r}=\nicefrac{\mathbb{F}_{2}\left[x_{1},x_{2},\cdots,x_{r}\right]}{\left(x_{1}^{l_{1}}-1,x_{2}^{l_{2}}-1,\cdots,x_{r}^{l_{r}}-1\right).}$ Here, the quotient ring $R_{r}$ has a basis consisting of the monomials
$x_{1}^{i_{1}}x_{2}^{i_{2}}\cdots x_{r}^{i_{r}}$, where $0\leq i_{1}\leq l_{1}-1,0\leq i_{2}\leq l_{2}-1,\cdots,0\leq i_{r}\leq l_{r}-1.$
In this work, we define the variable $x_{i}$ with $1\leq i \leq r$, using the following
encoding:
\begin{enumerate}
\item Let $enc_\text{bin}(i)=b_{q}b_{q-1}\cdots b_{2}b_{1}$ be the binary
encoding of $i$ using $ q = \lfloor \log r \rfloor +1$ bits.
\item Let $t_{\alpha}\equiv S_{l_{i}}^{b_{\alpha}}$ where $S_{l_{i}}$
is a cyclic shift matrix of size $l_{i}\times l_{i}$.
\item $x_{i} \equiv t_{q}\otimes t_{q-1}\otimes\cdots t_{2}\otimes t_{1}.$
\end{enumerate}

Note that the above encoding is not unique and has been chosen for
its simplicity and one can trivially define other encodings. Moreover,
the different variables are not independent. For example, in the trivariate
case, the quotient ring is given by
\[
R_{3}^{\prime}=\nicefrac{\mathbb{F}_{2}\left[x_{1},x_{2},x_{3}\right]}{\left(x_{1}^{l_{1}}-1,x_{2}^{l_{2}}-1,x_{3}-x_{1}x_{2}\right),}
\]
which is a subset of 
\[
R_{3}=\nicefrac{\mathbb{F}_{2}\left[x_{1},x_{2},x_{3}\right]}{\left(x_{1}^{l_{1}}-1,x_{2}^{l_{2}}-1,x_{3}^{l_{3}}-1\right).}
\]
Thus, our choice of encoding leads to a reduction in the size of the set $R_r$. In general, however, one can trivially define variables such that they are all independent. One such candidate encoding could be one-hot encoding of $i$, defined as $enc_{\text{one-hot}}(i)=b_{q}b_{q-1}\cdots b_{2}b_{1}$ with $q=r$. For example, $enc_\text{one-hot}(4) = 01000$, for $r=5$. The step $2$ and $3$ remains same as the last encoding. Note that the BB code corresponds to $r=2$ case with either choice of encoding, one-hot or binary. This is because the aforementioned two encodings are same for $r=2$.

For concreteness, in this work, we will focus on $r=3$ with binary encoding, which we call TB codes. We start with the identity matrix $I_{m}$ and the cyclic shift matrix $S_{l}$ of size $l \times l$ with $(S_{l})_{kj} = \delta_{j, (k + 1)\,\text{mod}\,l}$ for integer-valued $l$ and $m$. %
Let now
\begin{equation*}
x\equiv x_1= S_{l} \otimes I_m\,,\hspace{0.2cm} y\equiv x_2 = I_{l} \otimes S_m\,, \hspace{0.2cm} z\equiv x_3 = S_{l} \otimes S_{m}
\end{equation*}
be matrices of dimension $(l \cdot m) \!\times\! (l \cdot m)$ that commute with each other $[x_i, x_j] = 0$ $ \forall\,i,j \!\in\! \{1,2,3\}$. Now, we define the pair of matrices
\begin{equation}
    A = \sum^{\mathcal{W}_{A}}_{j=1} A_{j} \quad \text{and} \quad B = \sum^{\mathcal{W}_{B}}_{j=1} B_{j}
\end{equation}
generating a code of weight $\mathcal{W} = \mathcal{W}_{A} + \mathcal{W}_{B}$ with each matrix $A_{j}$ and ${B}_{j}$ being powers of $x, y$ or $z$. Working in the binary field $\mathbb{F}_{2} = \{0, 1\}$, we ensure avoiding cancellation of terms by only allowing for unique terms in both $A$ and $B$.
In order to fully describe a TB code, we need to further define the stabilizer check matrices
\[ H_{X} = [A|B] \quad \text{and} \quad H_{Z} = [B^{T}|A^{T}] \]
Following this definition, we can then define $X$-type and $Z$-type check operators being a row $v \!\in\! \mathbb{F}_{2}$ of $H_{X}$ and $H_{Z}$, respectively. Since $[A, B] = 0$, these satisfy the stabilizer code condition requiring all $X$-type checks to commute with all $Z$-type checks, translating to $H_{X} H_{Z}^{T} = 0$. We emphasize that the $A_{j}$ and ${B}_{j}$ terms can be understood as introducing edges between checks and vertices on the Tanner Graph of the TB code, one per term. This viewpoint is crucial in understanding physical implementations in the Code Layout section. Further details are provided in Sec.~D of the \SMLong{} (\SM{})~\cite{lukassimtobiaskishorSuppMaterial}.

We then find the TB code with parameters $\dsl n, k, d\dsr$ in accordance with Lemma 1 of Ref.~\cite{Bravyi_2024}
\begin{equation}
    \begin{split}
        n &= 2lm, \quad k = 2 \cdot \text{dim}(\text{ker}(A) \cap \text{ker}(B)) \\ d &= \text{min}\{|v|: v \in \text{ker}(H_{X}) \backslash \text{rs}(H_{Z}) \}
    \end{split}
\end{equation}
with $|v|$ as the Hamming weight of vector $v$.

\begin{table}[t!]
    \centering
    \setlength{\tabcolsep}{4pt} %
    \begin{tabular}{ccccccc}
        \toprule
        \( \dsl n, k, d\dsr \) & $\mathcal{W}$ & \( k/n \) & \( r_{\text{TB}}/r_{\text{SC}} \) & toric & bi-planar & $N_\text{L}$  \\
        \midrule
        $\dsl 144, 2, 12\dsr$ & 4 & $1/72$ & 2.0 & $\times$ & $\checkmark$ & $1$ \\
        $\dsl 30, 4, 5\dsr$ & 5 & $1/7$ & 3.3 & $\checkmark$ & $\checkmark$ & $7$ \\
        $\dsl 30, 6, 4\dsr$ & 6 & $1/5$ & 3.2 & $\checkmark$ & $\checkmark$ & $4$ \\
        $\dsl 30, 4, 5\dsr$ & 7 & $1/7$ & 3.3 & $\checkmark$ & ? & $4$ \\
        \bottomrule
    \end{tabular}
    \caption{Best TB-QLDPC codes that we found ordered by weight $\mathcal{W}$ of the stabilizer checks.  The code parameters $\dsl n, k , d\dsr$ correspond to $k$ logical qubits, $n$ physical qubits and code distance $d$. The encoding rate $r=k/n$ is rounded to the next smallest integer value. 
    In column $r_{\text{TB}}/r_{\text{SC}}$, we show the improvement factor in encoding rate compared to a surface code with the same code distance and number of logical qubits. Last columns indicate whether code has toric structure, can be encoded on a bi-planar lattice, and the number $N_\text{L}$ of logical circuits implementable by transversal operations and SWAPs found via automorphism groups.   %
    }
    \label{tab:ldpc_poster_codes}
\end{table}

\prlsection{Results}\label{sec::mainresults} 
In Table~\ref{tab:ldpc_poster_codes}, we provide an overview of our TB codes of weights four to seven. For a given distance $d$ and number of logical qubits $k$, our codes require less physical qubits than surface codes  which need $n=kd^{2}$ physical qubits~\cite{Bravyi_1998, Kitaev_2003}. %
\revA{For our codes, we also find $N_\text{L}$ fault-tolerant logical circuits, which can be implemented via transversal operation and SWAPs, via automorphism groups~\cite{sayginel2024fault} (see \SM{}~\cite{lukassimtobiaskishorSuppMaterial} for explicit circuits).}
Note that we found even more TB codes of weight four to seven with favourable properties, which are shown in Sec.~A in the \SM{}~\cite{lukassimtobiaskishorSuppMaterial}.

To characterize the error capabilities of our codes, we study the pseudo threshold $p_{0}$. It is defined as the value satisfying the break-even relation $p_\text{L}(p) = p_\text{unenc}(p)$ with $p_\text{unenc}(p)$ being the probability that at least one unencoded qubit suffers an error. 
We also fit the logical error $p_\text{L}$ with the heuristic formula~\cite{bravyi2013simulation, Bravyi_2024}
\begin{equation}
    p_\text{L}=p^{ d_\text{fit}/2 }\exp(c_0+c_1p+c_2p^2)
\label{eq:heuristic_fit}
\end{equation}
where for surface codes and bicycle codes it has been observed that the fitted distance corresponds to the actual distance via $d_\text{fit}/2=\lceil d/2 \rceil$~\cite{bravyi2013simulation}. %

\begin{figure}[!h]
    \centering
\includegraphics[width=0.35\textwidth]{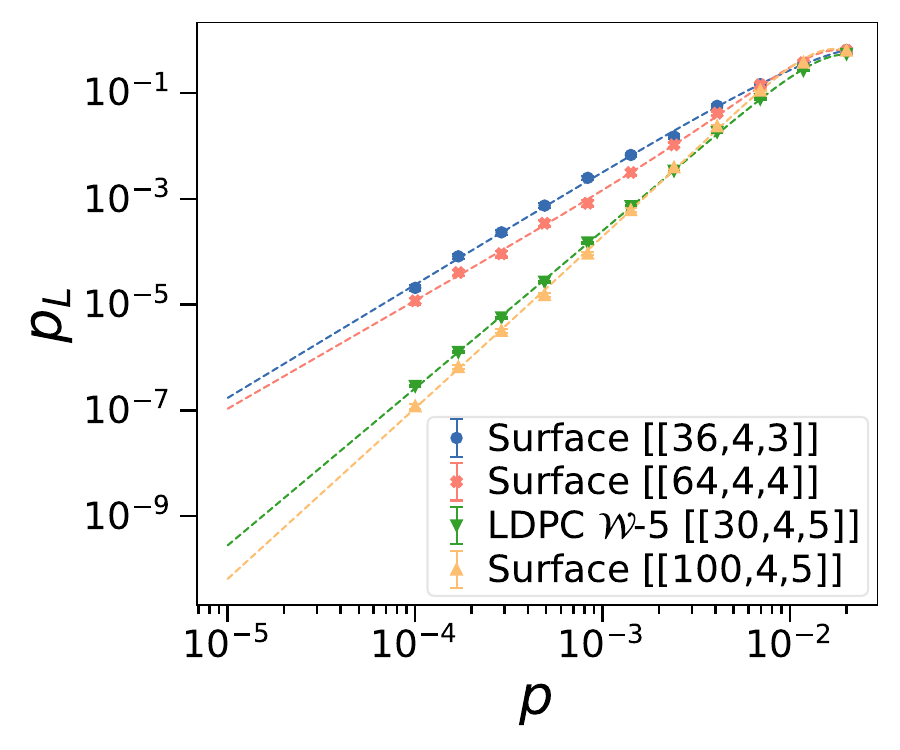}
    \caption{\revA{Circuit-level noise for our weight-5 $\dsl 30, 4, 5\dsr$ TB code and surface codes. We show logical error $p_\text{L}$ against physical error rate $p$. Using~\eqref{eq:heuristic_fit}, we find $d_\text{fit}=\{4.3, 4.1, 6.0, 6.4\}$, and pseudo-threshold $p_0=\{1.2\cdot 10^{-3},2.2\cdot 10^{-3},4.0\cdot 10^{-3},3.6\cdot 10^{-3}\}$.}}
    \label{fig:circuitW5}
\end{figure}

\revA{We simulate our codes under circuit-level depolarizing noise, where we explicitly simulate the syndrome measurement circuit including one-qubit, two-qubit, measurement, state preparation and idling noise following Ref.~\cite{Bravyi_2024} (see Appendix~\ref{app:circuitlevel}). To decode, a linearized noise model is constructed by simulating each error individually and constructing an effective parity check matrix, which neglects cancellations between errors~\cite{Bravyi_2024}. To decode, we use the Ordered Statistics postprocessing step Decoder (BP-OSD) of Refs.~\cite{roffe2020decoding,panteleev2021degenerate} where we find the most likely logical error given the measured syndrome.

We study our weight-5 $\dsl 30, 4, 5\dsr$ TB code under circuit-level depolarizing noise in Fig.~\ref{fig:circuitW5}. As comparison, we show surface codes with the same number of logical qubits. We find that our code has the same scaling with $p$ as a surface code with more than three times the number of qubits. }

\begin{figure}[!h]
    \centering
    \includegraphics[width=0.35\textwidth]{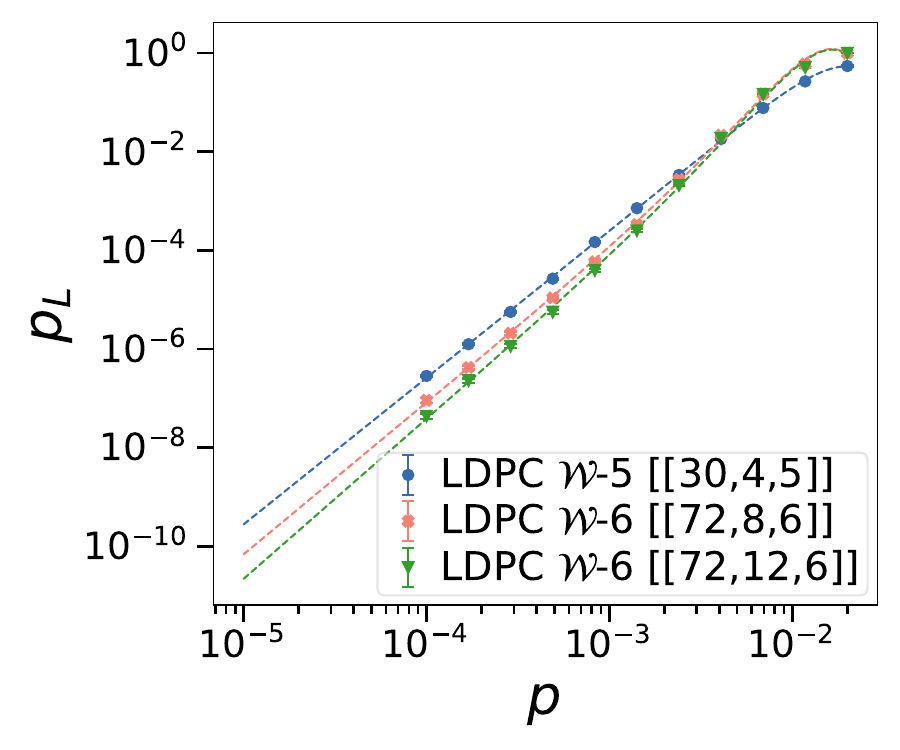}
    \caption{\revA{Circuit-level noise for  our weight-5 $\dsl 30, 4, 5\dsr$ TB code (blue), the previously proposed weight-6 $\dsl 72,8,6\dsr$ BB code of Ref.~\cite{Berthusen_2024} (orange), and weight-6 $\dsl 72,12,6\dsr$ BB code of Ref.~\cite{Bravyi_2024} (green). Dashed line fit with~\eqref{eq:heuristic_fit}, where we find $d_\text{fit}=\{6.0, 6.1, 6.5\}$, and pseudo-threshold $p_0=\{4.0\cdot 10^{-3}, 5.0\cdot 10^{-3},6.0\cdot 10^{-3}\}$.}} 
    \label{fig:comparison_IBM_Maryland}
\end{figure}

\revA{In Fig.~\ref{fig:comparison_IBM_Maryland}, we compare the noise suppression characteristics of our weight-$5$ $\dsl 30,4,5\dsr$ TB code (Fig.~\ref{fig:sketch}) with the weight-6 $\dsl 72,12,6\dsr$ code of Ref.~\cite{Bravyi_2024} and the weight-6 $\dsl 72,8,6\dsr$ code of Ref.~\cite{Berthusen_2024}. We find similar noise performance and scaling for all three codes, where notably our code requires less than half of qubits to implement. %
We show additional results on the performance of our codes in Sec.~B, C of the \SM{}.}

\prlsection{Code Layout}
\label{section:codelayout}
We now discuss the physical layout needed to implement our TB-QLDPC codes where we consider bi-planarity and toric layout. Bi-planarity means that the Tanner Graph has thickness-2, where thickness refers to the minimum number of separate two-dimensional planar layers necessary to embed the graph. Operationally, thickness tells us how many circuit boards are needed in a physical construction of the QLDPC code. 

\begin{proposition}[Bi-planar Architecture] 
All TB-QLDPC codes of weight-4, where $A = A_1 + A_3$, B = $B_1 + B_3$, all codes of weight-5 such that A = $A_1 + A_3$, B = $B_1 + B_2 + B_3$, and all codes of weight-6 such that  A = $A_1 + A_2 + A_3$, B = $B_1 + B_2 + B_3$ (the case presented in Ref.~\cite{Bravyi_2024}), or A = $A_1 + A_3$, B = $B_1 + B_2 + B_3 + B_4$ allow for a bi-planar architecture of thickness $\theta = 2$. The bi-planar decomposition can be computed in time $O(n)$.
\end{proposition}
The proof is outlined in Proposition 6 in \SM{}~\cite{lukassimtobiaskishorSuppMaterial}. 
It implies that our weight $4$ to weight $6$ codes shown in Table~\ref{tab:ldpc_poster_codes} can be implemented in a bi-planar lattice architecture (see Fig.~\ref{fig:sketch}), which makes them suitable candidates for near-term hardware implementations
It is currently unknown whether the weight-7 code is bi-planar. Beyond bi-planarity, we use the concept of a toric layout to simplify the design process in physical implementations as elaborated in Proposition 5 (\SM{}~\cite{lukassimtobiaskishorSuppMaterial}). We view the Tanner Graph as embedded on a torus, allowing for edge crossings, see Fig.~\ref{fig:sketch}. 
\begin{definition}[Toric Layout~\cite{Bravyi_2024}]
\label{definition:toriclayout}
A TB-QLDPC code has a toric layout $\iff \exists$ positive integers $\mu, \lambda$ such that its Tanner Graph has a spanning sub-graph isomorphic to the Cayley Graph of $\mathbb{Z}_{2\mu} \times \mathbb{Z}_{2\lambda}$. 
The parameters $2\mu$ and $2\lambda$ determine the width and length of the toric layout for the MB code's Tanner Graph. 
\end{definition}

Using the proposition below, we obtain an easy check for whether a TB code has a toric layout as further explained in SM~\cite{lukassimtobiaskishorSuppMaterial}. We first define M to be the set of monomials $M = \{x^i y^j \, \vert \, i = 0,1,.., l-1, j = 0,1, \dots,m-1\}$.

\begin{proposition}[TB-QLDPC Toric Layout Criterion for arbitrary weight, generalised from Ref.~\cite{Bravyi_2024}] 
\label{proposition:toriclayout}
    A TB-QLDPC code (\text{QC}($A$, $B$)) of weight $\mathcal{W}=\mathcal{W}_A+\mathcal{W}_B$ has a toric layout $\iff \exists \, i, j \in \{1,..,\mathcal{W}_A\}, g, h \in \{1,..,\mathcal{W}_B\}$ such that 
\begin{enumerate}
    \item \(\langle A_i A_j^T, B_g B_h^T \rangle = M\) and
    \item \(\operatorname{ord}(A_i A_j^T) \operatorname{ord}(B_g B_h^T) = lm\).
\end{enumerate}
\end{proposition}

Although Ref.~\cite{Bravyi_2024} proves Proposition~\ref{proposition:toriclayout} for the weight-6 case where $A = A_1 + A_2 + A_3$, $B = B_1 + B_2 + B_3$, we emphasize that the proposition holds more generally because none of their argument requires that  $i, j, g, h \in \{1,2,3\}$. In fact, we may carry out an identical argument for generic weights $A = \sum^{\mathcal{W}_{A}}_{j=1} A_{j}$ and $ B = \sum^{\mathcal{W}_{B}}_{j=1} B_{j}$. As an example of the generality of Proposition~\ref{proposition:toriclayout}, our weight-7 $\dsl 30, 4, 5\dsr$ code was numerically confirmed to have a toric layout, see Sec.~A in the \SM{}~\cite{lukassimtobiaskishorSuppMaterial}.

Our codes have $r=2$ and thus according to our toric layout criterion can be realized on a 2-dimensional Euclidean space, which is directly accessible for current experiments. In general, for  multivariate group algebra of $r$ independent variables, we obtain an analogous toric layout definition and criterion where the code can be realized on a $r$-dimensional torus for $\mathbb{Z}_{2\mu_1}\times  \dots\times \mathbb{Z}_{2\mu_r}$. This higher dimensional toric layout arises because each independent variable introduces an independent orthogonal direction in Euclidean space.

\begin{proposition}[TB-QLDPC Toric Layout Edge Translational Invariance]

For any TB-QLDPC code (with Tanner Graph $G$) of sparsity $\mathcal{W}$ with a toric layout, for any vertex $v_T$ of a fixed type $T \in \{L, R, X, Z\}$, all of the $\mathcal{W}$ edges of $v_T$, including the long-range edges, have translationally invariant interaction vectors on the toric layout. 
\end{proposition}

\begin{figure}
    \centering
    \includegraphics[width=0.4\textwidth]{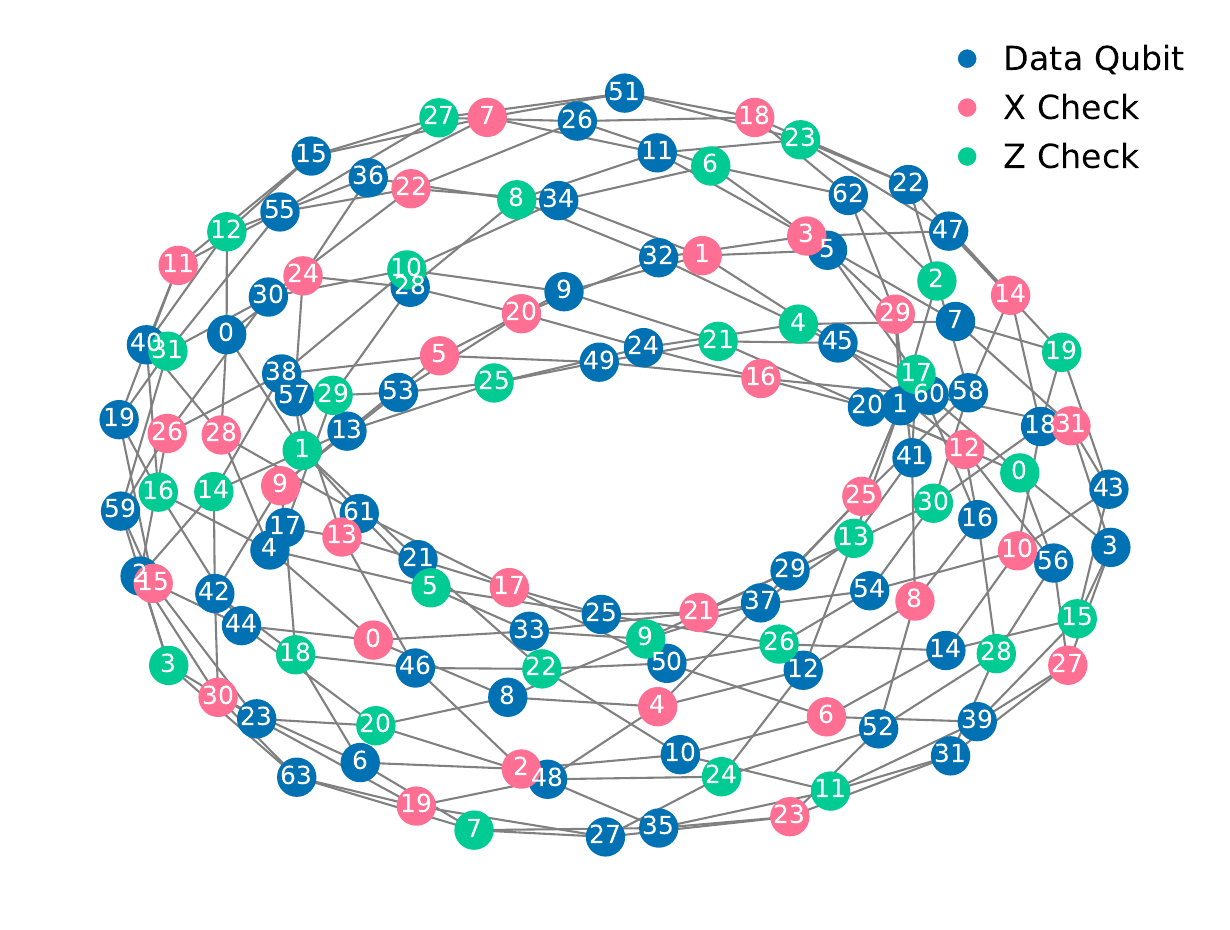}
    \caption{Tanner Graph of the weight-4 $\dsl 64, 2, 8\dsr$ code, as a tangled torus with layout parameters $\mu = l = 8, \lambda = m = 4$. The usual torus topology is disrupted by a tangle on the right side of the figure.}
    \label{fig:weight4_twistedtorus}
\end{figure}

For a weight-$\mathcal{W}$ TB code, four interactions remain local, while $\mathcal{W}-4$ interactions are non-local with translational symmetry. An example is shown in Fig.~\ref{fig:sketch}. This is facilitated by the cyclic property of $H_X$ and $H_Z$, as elaborated in the \SM{}~\cite{lukassimtobiaskishorSuppMaterial}. An analogous proposition holds for MB codes with a toric layout as well, by repeating the arguments in higher spatial dimensions. Using the test in Proposition~\ref{proposition:toriclayout}, we further conclude that all the weight-4 codes do not have a toric layout. However, they can possess a tangled toric layout, such as in Fig~\ref{fig:weight4_twistedtorus} by using the tool from Ref.~\cite{hagberg2008exploring}.

Intuitively, a tangled toric layout means that the code behaves as a surface code in the bulk, but with long-range edges produced by modifying the PBC along the horizontal and vertical edges of the square grid. \revA{We find that this gives a factor $2$ improved encoding rate compared to regular toric codes. Yet, we believe the tangled layout can be easier to implement than bi-planar layout for some  experimental platforms, and thus could be an interesting experimental direction.}
In the Bravyi-Poulin-Terhal (BPT) bound $kd^2 \leq cn$~\cite{Bravyi_2010}, for the toric code the locality constant $c = 1$ saturates the bound, but in our weight-4 $\dsl 64, 2, 8\dsr$ code we would need $c=2$, verifying its enhanced performance. We provide an $O(n)$ algorithm to determine the tangling parameters ($\sigma$, $\tau$). This algorithm applies more generally to two-block codes, defined as CSS QLDPC codes such that $H_X = [A|B]$ and $H_Z = [B^T |A^T ]$, where A and B commute and have the same size.

\begin{proposition}[Algorithm for Weight-4 ($\sigma$, $\tau$)-Tangled Toric Layout Parameters]
\label{proposition:weight4_tangled_toric_layout}
For any two-block CSS QLDPC code of sparsity 4 with equal X and Z degree per data qubit (2 each), suppose it has a tangled toric layout known to have torus parameters $\mu$,$\lambda$.
Then there is an $O(n)$ time algorithm to find $\sigma$, $\tau$.
\end{proposition}

The key idea of the algorithm is to embed a spanning subgraph onto the rectangular 2-D grid of size $2\mu \times 2\lambda$. The unassigned edges on the grid give us $\sigma$ and $\tau$ respectively.

\prlsection{Discussion}\label{sec::conclusions}
In this work, we introduce MB-QLDPC codes as multi-variable generalisation of BB-QLDPC codes, which have only two variables.
We study TB-QLDPC with three variables in detail and present compact codes with high encoding rates and strong noise suppression. 
Moreover, by optimised search we find codes with more resource-efficient parameters than corresponding surface codes. We obtain promising codes with only weights four and five, which could offer a more hardware-friendly implementation compared to recently found weight-6 BB-QLDPC codes from Ref.~\cite{Bravyi_2024, Berthusen_2024,Scruby_2024_singleshotdecodable}. For example, we find a highly compact weight-5 $\dsl 30, 4, 5\dsr$ code which has a toric bi-planar layout and a single long-range connection, which shows promise for practical implementations on current hardware. \revA{In particular, our code can be implemented with $60$ physical qubits with comparable performance to surface codes of $200$ qubits, as well as weight-6 codes of Refs.~\cite{Bravyi_2024,Berthusen_2024}. Thus, our code is especially promising when the qubit number is limited, as is the case with current quantum computers. 
We also found a highly compact depth-7 syndrome measurement circuit for weight-5 codes, which we find preserves the distance even under circuit-level noise and could be relevant for other weight-5 codes. We also find $7$ fault-tolerant logical operations via automorphism groups, which can be implemented transversally combined with SWAPs~\cite{sayginel2024fault}.
We highlight that our code  is easier to implement in practice as it has only weight-$5$ parity checks with only one long-range connection, compared to weight-$6$ checks with two long-range connections~\cite{Bravyi_2024,Berthusen_2024}. As long-range connection is a major bottleneck in superconducting qubit setup requiring ancillas or long-range couplers~\cite{Berthusen_2024}, our code essentially reduces the cost of this problem in half.
}

Further, we make progress on the properties of TB-QLDPC codes: We show that up to weight 6, all TB codes have a bi-planar structure and further provide methods to check for a toric layout, which could be used to systematically search for codes with experimental-friendly layout, significantly reducing the search space of codes.
Future research could combine our MB-QLDPC codes with subsystem codes~\cite{poulin2005stabilizer,bravyi2012subsystem} and Floquet codes~\cite{hastings2021dynamically,gidney2021fault,vuillot2021planar,haah2022boundaries,higgott2023constructions,fahimniya2023fault} for subsystem MB bicycle codes and Floquet MB bicycle codes which could reduce the weight of check operators. 
A potential method to construct such codes is through Floquetification via ZX calculus~\cite{bombin2023unifying,townsend2023floquetifying}.
In the future, automating the discovery of MB-QLDPC with favourable parameters, such as weight, Tanner graph thickness, encoding rate, and distance, could be achieved using machine learning techniques like reinforcement learning. Alternatively, one could study the structure of polynomials analytically to directly find MB codes with favorable parameters and layout. Finally, it would be interesting to develop adaptive MB-QLDPC codes using quantum combs~\cite{tanggara2024strategic}.
Finally, it would be interesting to find efficient methods  to check whether a code is bi-planar, which  could be done by leveraging the symmetries in the Tanner Graph of QLDPC codes such as its toric layout or translation invariance. %

\begin{acknowledgments}
\prlsection{Acknowledgments}
We thank Sergey Bravyi and Patrick Rall for a fruitful discussion in the beginning of the project, and Victor Albert, Arne Løhre Grimsmo and Daniel Spencer for helpful comments. LV acknowledges the financial support by the Baden-Württemberg Foundation and Ernst-Wilken Foundation for his research stay at CQT. This research is supported by A*STAR C230917003.
\end{acknowledgments}

\onecolumngrid

\clearpage
\begin{center}
\textbf{\large Appendix}
\end{center}

\section{Background}\label{app:background}
QEC codes are essential for executing high-fidelity computations on noisy quantum computers. Without QEC, errors accumulate during circuit execution, yielding an unreliable output. 
Stabilizer codes~\cite{Gottesman_1997, Calderbank_1997} represent a class of QEC codes defined by their stabilizer group, which is an abelian subgroup of the Pauli group on $n$ qubits that keeps the codespace invariant. The codespace of a stabilizer code is equivalently described as the joint $+1$-eigenspace of the stabilizer generators $S = \langle S_{1}, S_{2}, \dots, S_{r} \rangle$. For a $\dsl n, k, d\dsr$ code with $n$ physical qubits, $k$ logical qubits, and distance $d$, the number of linearly independent generators is $r = n\!-\!k$. A stabilizer code qualifies as a QLDPC code if each qubit supports a constant number of stabilizer generators and each generator has a constant weight independent of $n$. A stabilizer code is classified as a CSS code~\cite{Calderbank_1996, Steane_1996} if each generator is a tensor product of $X$ and $I$ or a tensor product of $Z$ and $I$. Although the surface code is a QLDPC code, its encoding rate $k/n$ approaches zero as $n$ approaches infinity, contributing to its high overhead. In contrast, alternative QLDPC codes maintain asymptotically constant encoding rates while preserving or improving the $d=\mathcal{O}(\sqrt{n})$ distance scaling of the surface code~\cite{Tillich_2013, Hastings_2021, Breuckmann_2021, Panteleev_2021, Panteleev_2022, Leverrier_2022, Dinur_2023}.

Based on its stabilizer description, one can construct the Tanner graph $T(S) = (V_{q} \!\sqcup\! V_{S}, E)$ of a QEC code. This graph includes a vertex $q \!\in\! V_{q}$ for each data qubit and a vertex $s \!\in\! V_{S}$ for each stabilizer generator. An edge $(q, s) \!\in\! E$ exists between vertices $q \!\in\! V_{q}$ and $s \!\in\! V_{S}$ if the generator $s$ acts non-trivially on qubit $q$. The Tanner graph of a QLDPC code has a degree bounded by the constant $\mathcal{W}$, called the sparsity of the graph.

To check whether the encoded quantum information has left the logical subspace, the stabilizer generators are measured. Assuming the $n$ data qubits are in a code state, we expect a $+1$ outcome for all stabilizer generator measurements. A $-1$ outcome signals an error that anticommutes with the corresponding generator. These measurement outcomes form a classical syndrome, which is then used as input for a decoding algorithm to correct the errors while preserving the original logical state.

For CSS codes, the stabilizer generators either all involve measurements in $X$ or $Z$ basis. Thus, the decoding problem can be done separately for $x$ and $z$ errors. For example,  the feasible $z$ error configurations $e_\text{z}$ are given by $s_\text{x}=H_\text{x} e_\text{z}$, where $H_\text{x}$ is the parity check matrix for $X$ stabilizers and $s_\text{x}$ the syndrome outcomes. One can similarly define for $x$ errors $s_\text{z}=H_\text{z} e_\text{x}$.
By determining the correct $e_\text{z}$ configuration, we can apply correction operations to correct the error. As long as the number of errors is smaller than $\lfloor d/2\rfloor$, one can always find a correction operation that does not introduce a logical error.
Given $s_\text{x}=H_\text{x} e_\text{z}$, there are multiple feasible error configurations $e_\text{z}$. Decoding aims to find the error that has most likely occurred, e.g. the error with the lowest weight. There are many different kinds of decoders with varying speed and accuracy, such as decoders based on maximum-likelihood~\cite{bravyi2014efficient}, tensor-networks~\cite{ferris2014tensor}, minimum-weight perfect matching~\cite{dennis2002topological}, neural-networks~\cite{torlai2017neural} or union-finding~\cite{delfosse2021almost}.
The popular choice for surface codes, minimum-weight perfect matching, requires that each error affects at most two stabilizer generators such that the decoding problem can be mapped onto a graph~\cite{higgott2023sparse}. For QLPDC codes this is not possible in general.
In this work, we use Belief propagation with the
Ordered Statistics postprocessing step Decoder (BP-OSD) proposed in Ref.~\cite{roffe2020decoding,panteleev2021degenerate}. 
Belief propagation is a heuristic message passing algorithm that estimates the single-bit marginals of a probability distribution.

\section{Circuit-level noise}\label{app:circuitlevel}
\revA{
For our simulation of the codes under circuit-level noise, we follow the approach of Ref.~\cite{Bravyi_2024}. We explicitly construct the syndrome read-out circuits, and include errors for one-qubit and two-qubit gates, idling noise, read-out errors and state preparation errors for the ancillas needed to measure the syndrome. 
We assume depolarizing noise with probability $p$, with $3$ possible errors for single-qubit gates (and idling errors), measurement and state preparation errors, as well as $15$ different errors possible for two-qubit gates.
For the simulation, we repeat the syndrome measurement $d$ times, then perform the decoding, and apply a correction on the logical operators. If any logical operator now has flipped, we say a logical error has occurred. 
We estimate the logical error per syndrome measurement cycle $p_\text{L}$ via $p_\text{L}=1-(1-p_{d,\text{L}})^{1/d}$, where $p_{d,\text{L}}$ is the logical error probability after $d$ cycles.

The circuit for the syndrome measurements can be arranged in different manners which can strongly affect the logical error and circuit-level distance $d_\text{circ}$~\cite{Bravyi_2024}. 
We upper bound $d_\text{circ}$ by a linearized error model which is then solved using BP-OSD~\cite{Bravyi_2024}. Here, errors for $x$ and $z$ checks can be decoded separately.

A key challenge is to design compact syndrome measurement circuits while maintaining a high  circuit-level distance $d_\text{circ}\leq d$ while making sure that the syndrome circuit leaves the logical operators unchanged.
For weight-6 BB codes, a syndrome measurement circuit of depth 8 was found in Ref.~\cite{Bravyi_2024}, in contrast to depth-12 %
for a naive sequential measurement of X and Z syndromes. 

Here, we design such compact syndrome measurement circuits for our weight-5 $\dsl 30, 4, 5\dsr$ TB code. 
A naive sequential syndrome measurement yields a depth-10 circuit with $d_\text{circ}\leq4$. One can adapt the syndrome readout of Ref.~\cite{Bravyi_2024} of weight-6 for weight-5 by leaving out the sixth check. However, we find this yields a reduced circuit level distance $d_\text{circ}\leq4$ for our weight-5 $\dsl 30, 4, 5\dsr$ code. 
We searched for better constructions which cumulated in our Fig.~\ref{fig:sketch}b which using extensive numerics yields $d_\text{circ}\leq5$, which we believe to be tight. We further discuss the circuit construction in SM~\cite{lukassimtobiaskishorSuppMaterial} G.
We show the performance of the different syndrome measurement circuits in Fig.~\ref{fig:circuittypes}, finding that our circuit has better noise suppression and scaling in $p$.

\begin{figure}[htbp]
	\centering	\subfigimg[width=0.50\textwidth]{a}{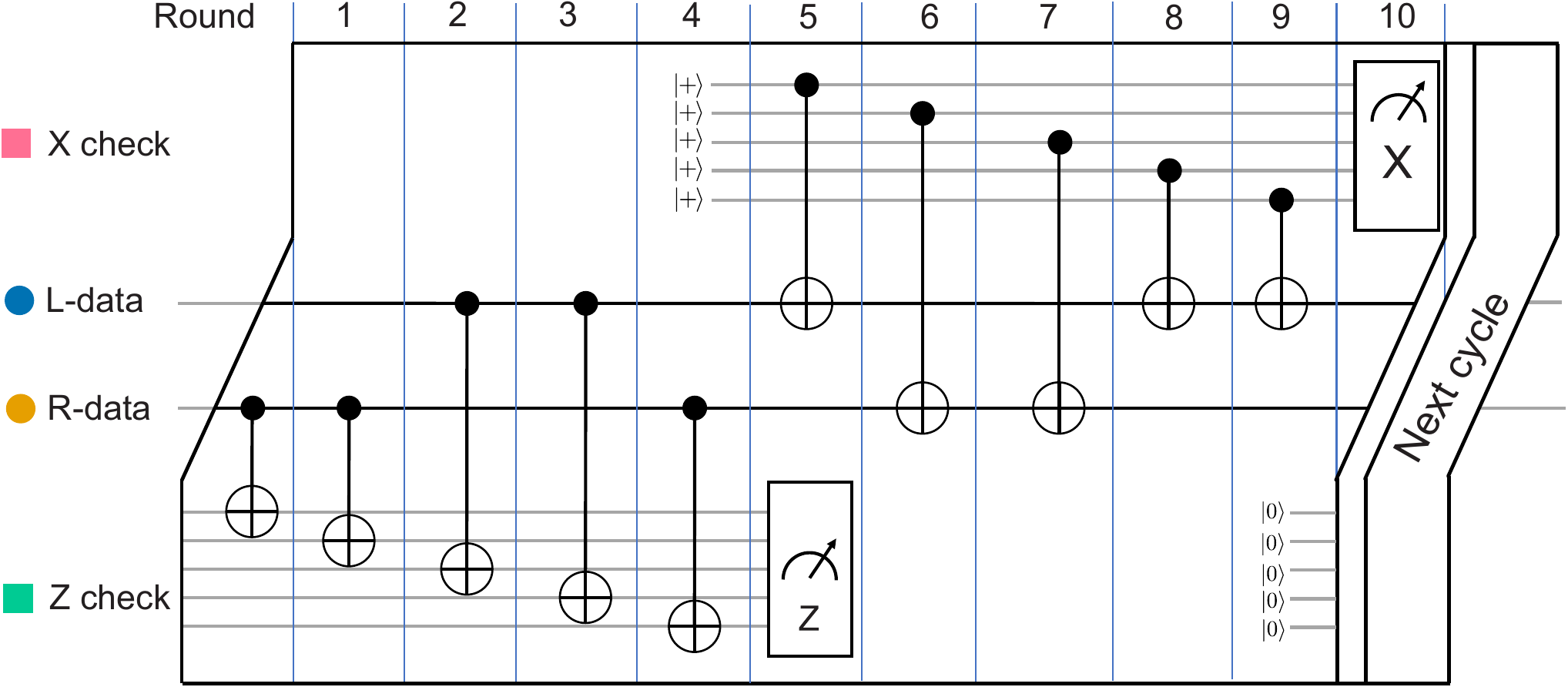}
 \subfigimg[width=0.39\textwidth]{b}{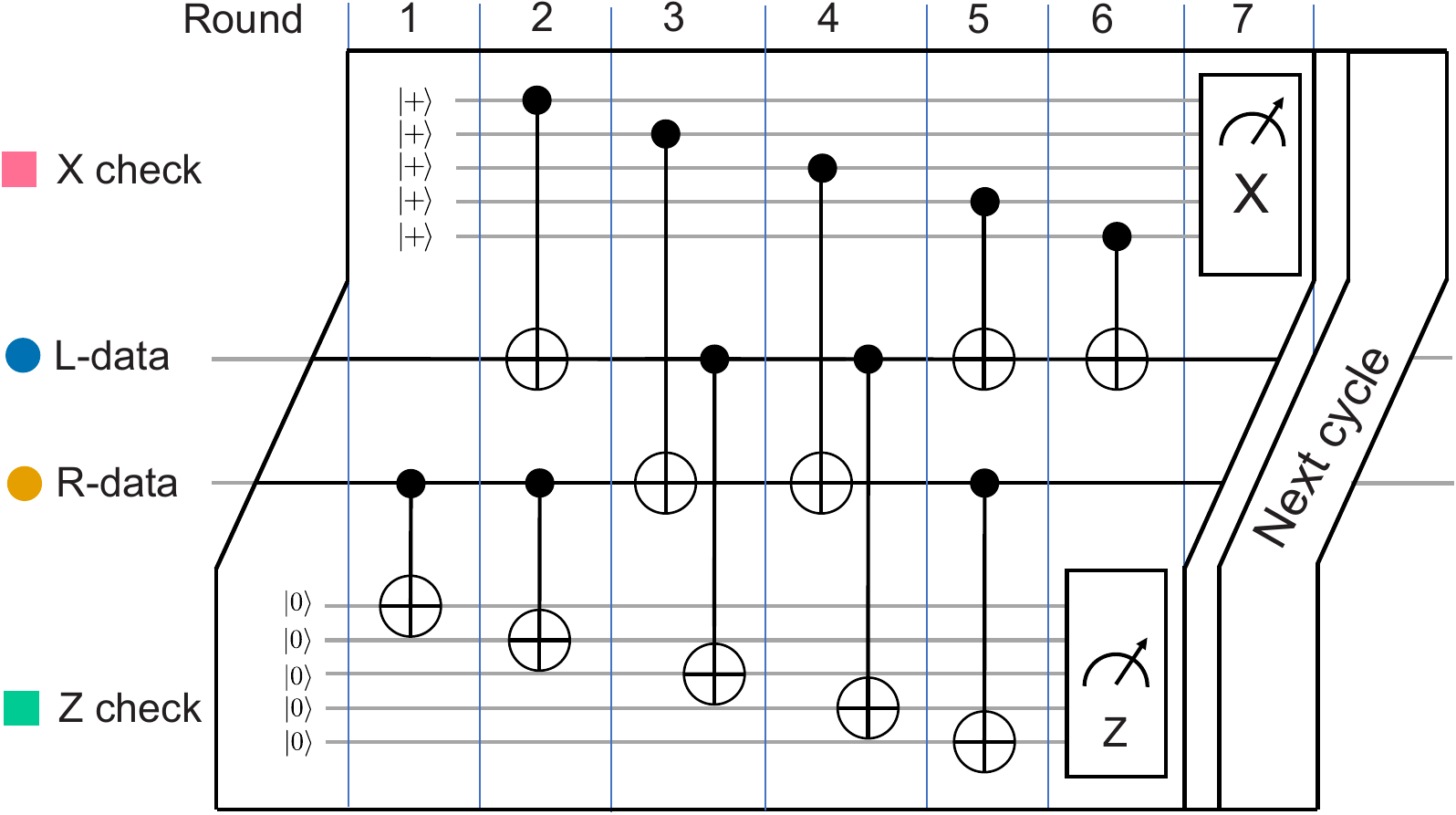}
	\caption{Additional types of syndrome circuits for weight-5 BB codes. We have \idg{a} sequential measurement of syndrome X and Z circuit of depth 10 \idg{b} interleaved circuit adapted from weight-6 BB codes adapted from Ref.~\cite{Bravyi_2024} by simply leaving out the sixth check.
    For our weight-5 $\dsl 30, 4, 5\dsr$, both circuits have only  circuit-level distance $d_\text{circ}\leq4$, while our circuit of Fig.~\ref{fig:sketch}b has $d_\text{circ}\leq5$.
	}
	\label{fig:circuittypes}
\end{figure}

In Fig.~\ref{fig:compsyndromW5}, we compare the circuit-level logical error $p_\text{L}$ against physical error $p$ for different syndrome measurement schemes for our weight-5 $\dsl 30, 4, 5\dsr$ code. The fit demonstrates the superior scaling of the logical error for the distance-preserving syndrome measurement circuit of Fig.~\ref{fig:sketch}b compared to the other circuits.}

\begin{figure}[!h]
\centering
\includegraphics[width=0.3\textwidth]{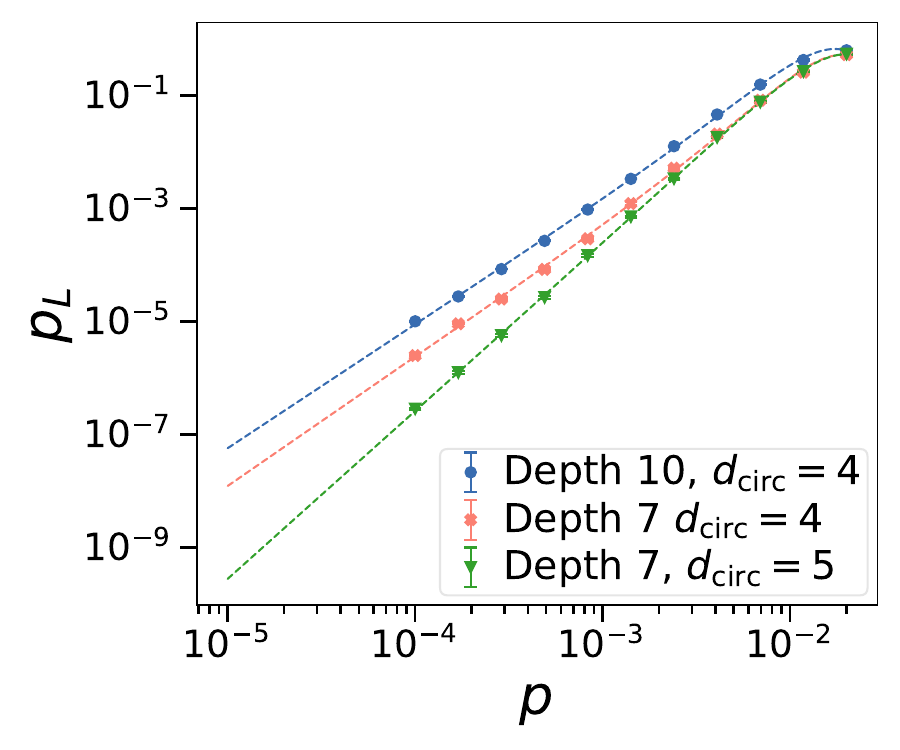}
    \caption{Comparison different syndrome readout for circuit-level noise: We show logical error $p_\text{L}$ against physical error rate $p$ for our weight-5 $\dsl 30, 4, 5\dsr$ code. Blue curve shows sequential syndrome measurement circuit of Fig.~\ref{fig:circuittypes}a with depth 10 and $d_\text{circ}=4$, orange Fig.~\ref{fig:circuittypes}b with depth 7 and $d_\text{circ}=4$, and green Fig.~\ref{fig:sketch}b of the main text with depth 7 and $d_\text{circ}=5$.
    The fitted distance is $d=\{4.4, 4.6, 6.0\}$s, and pseudothreshold $p_0=\{2\cdot10^{-3},3.6\cdot10^{-3},4.0\cdot10^{-3}\}$. } 
    \label{fig:compsyndromW5}
\end{figure}

\bibliography{references.bib}

\let\addcontentsline\oldaddcontentsline

\onecolumngrid

\newpage
\appendix

\clearpage
\begin{center}

\textbf{\large \SMLong{}}
\end{center}

We provide proofs and additional details supporting the claims in the main text.

\makeatletter
\@starttoc{toc}

\makeatother

\section{Codes overview}\label{app::codes}

\begin{table*}[!htbp]
    \centering
        \centering
        \renewcommand{\arraystretch}{1.2} %
        \begin{tabular*}{\textwidth}{@{\extracolsep{\fill}} lcccclccccccc}
        \multicolumn{10}{l}{Weight-4 QLDPC Codes} \\
        \toprule
        \( \dsl n, k, d\dsr \) & $r$ & \( l, m \) & \( r_{\text{BB}}/r_{\text{SC}} \) & $p_{0}$ & \( p_{L}(10^{-4}) \) & $A$ & $B$ & \(i, j, g, h\) & \(\mu, \lambda\) & toric & bi-planar \\
        \midrule
        $\dsl 112, 8, 5\dsr$ & $1/14$ & 7, 8 & 1.8 & 0.0298 & $2\!\times\!10^{-9}$ & $z^2 + z^6$ & $x + x^6$ & $\times$ & $\times$ & \(\times\) & \(\checkmark\) \\
        $\dsl 64, 2, 8\dsr$ & $1/32$ & 8, 4 & 2.0 & 0.0767 & $4\!\times\!10^{-13}$ & $x + x^{2}$ & $x^3 + y$ & $\times$ & $\times$ & \(\times\) & \(\checkmark\) \\
        $\dsl 72, 2, 8\dsr$ & $1/36$ & 4, 9 & 1.8 & 0.0863 & $2\!\times\!10^{-13}$ & $x + y^{2}$ & $x^{2} + y^{2}$ & $\times$ & $\times$ & \(\times\) & \(\checkmark\) \\
        $\dsl 96, 2, 8\dsr$ & $1/48$ & 6, 8 & 1.3 & 0.0911 & $4\!\times\!10^{-16}$ & $x^5 + y^6$ & $z + z^4$ & $\times$ & $\times$ & \(\times\) & \(\checkmark\) \\
        $\dsl 112, 2, 10\dsr$ & $1/56$ & 7, 8 & 1.8 & 0.097 & $2\!\times\!10^{-16}$ & $z^6 + x^5$ & $z^{2} + y^5$ & $\times$ & $\times$ & \(\times\) & \(\checkmark\) \\
        $\dsl 144, 2, 12\dsr$ & $1/72$ & 8, 9 & 2.0 & 0.1017 & $4\!\times\!10^{-19}$ & $x^3 + y^7$ & $x + y^5$ & $\times$ & $\times$ & \(\times\) & \(\checkmark\) \\
        \bottomrule
        \end{tabular*}
\end{table*}
\begin{table*}[!htbp]
        \centering
        \renewcommand{\arraystretch}{1.2} %
        \begin{tabular*}{\textwidth}{@{\extracolsep{\fill}} lcccclccccccc}
            \multicolumn{6}{l}{Weight-5 QLDPC Codes} \\
            \toprule
            \( \dsl n, k, d\dsr \) & $r$ & \( l, m \) & \( r_{\text{BB}}/r_{\text{SC}} \) & $p_{0}$ & \( p_{L}(10^{-4}) \) & $A$ & $B$ & \(i, j, g, h\) & \(\mu, \lambda\) & toric & bi-planar \\
            \midrule
            $\dsl 30, 4, 5\dsr$ & $1/7$ & 3, 5 & 3.3 & 0.0437 & $6\!\times\!10^{-10}$ & $x + z^4$ & $x + y^2 + z^2$ & $(1, 2, 2, 3)$ & $(5, 3)$ & \(\checkmark\) & \(\checkmark\) \\
            $\dsl 72, 4, 8\dsr$ & $1/18$ & 4, 9 & 3.6 & 0.0785 & $8\!\times\!10^{-14}$ & $x + y^3$ & $x^2 + y + y^2$ & $\times$ & $\times$ & \(\times\) & \(\checkmark\) \\
            $\dsl 96, 4, 8\dsr$ & $1/24$ & 8, 6 & 2.7 & 0.0823 & $1\!\times\!10^{-13}$ & $x^6 + x^3$ & $z^5 + x^5 + y$ & $(1, 2, 1, 2)$ & $(8, 6)$ & \(\checkmark\) & \(\checkmark\) \\
            \bottomrule
        \end{tabular*}
        \label{tab:ldpc_codes_weight5}
\end{table*}
\begin{table*}[!htbp]
        \centering
        \renewcommand{\arraystretch}{1.2} %
        \begin{tabular*}{\textwidth}{@{\extracolsep{\fill}} lcccclccccccc}
            \multicolumn{6}{l}{Weight-6 QLDPC Codes} \\
            \toprule
            \( \dsl n, k, d\dsr \) & $r$ & \( l, m \) & \( r_{\text{BB}}/r_{\text{SC}} \) & $p_{0}$ & \( p_{L}(10^{-4}) \) & $A$ & $B$ & \(i, j, g, h\) & \(\mu, \lambda\) & toric & bi-planar \\
            \midrule
            $\dsl 30, 6, 4\dsr$ & $1/5$ & 5, 3 & 3.2 & 0.0234 & $3\!\times\!10^{-7}$ & $x^4 + z^3$ & $x^4 + x + z^4 + y$ & $(1, 2, 1, 3)$ & $(5, 3)$ & \(\checkmark\) & \(\checkmark\) \\
            $\dsl 48, 6, 6\dsr$ & $1/8$ & 4, 6 & 4.5 & 0.0495 & $2\!\times\!10^{-10}$ & $x^2 + y^4$ & $x^3 + z^3 + y^2 + y$ & $\times$ & $\times$ & \(\times\) & \(\checkmark\) \\
            $\dsl 40, 4, 6\dsr$ & $1/10$ & 4, 5 & 3.6 & 0.0588 & $7\!\times\!10^{-11}$ & $x^2 + y$ & $y^4 + y^2 + x^3 + x$ & $\times$ & $\times$ & \(\times\) & \(\checkmark\) \\
            $\dsl48, 4, 6\dsr$ & $1/12$ & 4, 6 & 3.0 & 0.0698 & $3\!\times\!10^{-11}$ & $x^3 + y^5$ & $x + z^5 + y^5 + y^2$ & $(1, 2, 3, 4)$ & $(12, 2)$ & \(\checkmark\) & \(\checkmark\) \\
            \bottomrule
        \end{tabular*}
        \label{tab:ldpc_codes_weight6}
\end{table*}
\begin{table*}[!htbp]
        \centering
        \renewcommand{\arraystretch}{1.2} %
        \begin{tabular*}{\textwidth}{@{\extracolsep{\fill}} lcccccccccccc}
            \multicolumn{6}{l}{Weight-7 QLDPC Code} \\
            \toprule
            \( \dsl n, k, d\dsr \) & $r$ & \( l, m \) & \( r_{\text{BB}}/r_{\text{SC}} \) & $p_{0}$ & \( p_{L}(10^{-4}) \) & $A$ & $B$ & \(i, j, g, h\) & \(\mu, \lambda\) & toric & bi-planar \\
            \midrule
            $\dsl 30, 4, 5\dsr$ & $1/7$ & 5, 3 & 3.3 & 0.0507 & $5\!\times\!10^{-10}$ & $x^4 + x^2$ & $x + x^2 + y + z^2 + z^3$ & $(1, 2, 2, 4)$ & $(5, 3)$ & \(\checkmark\) & \(\times\) \\
            \bottomrule
        \end{tabular*}
        \label{tab:ldpc_codes_weight7}
    \caption{Summary of QLDPC codes sorted by encoding rate $r=k/n$. For each code, we provide the improvement in qubit-efficiency $r_{\text{BB}}/r_{\text{SC}}$ compared to the surface code. We also show physical error-rate below which error correction (assuming noise-free syndrome measurement) becomes a net-gain also known as the pseudo-threshold $p_{0}$, as well as the logical error rate $p_\text{L}$ for physical error rate $p=10^{-4}$. We also show the respective code polynomials $A$ and $B$. $i, j, g, h$ refer to the indices chosen to satisfy a toric layout, with $\mu, \lambda$ as the torus parameters such that the torus is embedded on a 2$\mu \!\times\! 2\lambda$ grid with periodic boundary conditions. If no suitable indices could be found that allow for a toric layout, the column value is set to $\times$.}
    \label{app:weight4-7_codes}
\end{table*}

\clearpage
\newpage

\section{Compare codes with different number of logical qubits}
For benchmarking, we compare planar surface codes (which have number of logical qubits $k=1$) with our QLPDC codes that have similar distance $d$, but a larger number of logical qubits $k$. 
We now show how to properly compare codes of different number of logical qubits $k$ to surface codes.

In particular, the planar surface code of distance $d$ can encode one logical qubit into a $d \!\times\! d$ grid of physical qubits, leading to code parameters $\dsl d^{2}, 1, d\dsr$~\cite{Asfaw_2023}. To encode $k$ logical qubits, we implement $k$ instances of the planar surface code, resulting in a $\dsl kd^{2}, k, d\dsr$ code.
The logical error probability per syndrome cycle $p_{L}$ for $k$ logical qubits can be directly computed from the logical error of a single planar surface code $p_{L}(1)$ via~\cite{Bravyi_2024}
\begin{equation}
    p_{L}(k) = 1 - (1 - p_{L}(1))^{k}
\end{equation}
with $ p_{L}(k)$ as the logical error probability of the surface code encoding $k$ logical qubits.

\section{Showcase codes}
\begin{figure}[!h]
    \centering
    \includegraphics[width=0.33\textwidth]{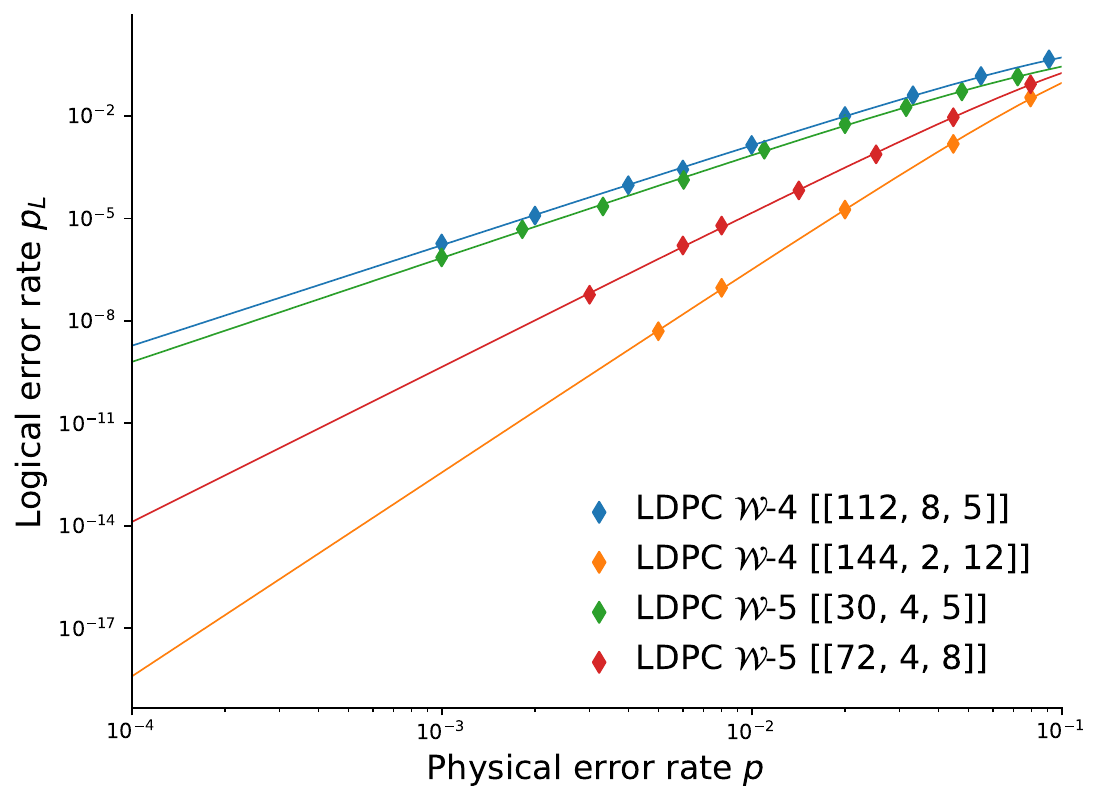}
    \caption{Comparison of noise properties among our TB codes of weight-$4$ and weight-$5$.
    We find the fitted code distances $d_\text{fit}=\{  5.9, 12.0, 6.1, 8.3\}$, respectively.
    }
    \label{fig:poster_codes_weight45}
\end{figure}

In Fig.~\ref{fig:poster_codes_weight45}, study codes of stabilizer weights four and five under local depolarizing noise, where we assume that the syndrome readout is noiseless. We performed decoding with the Belief propagation with the Ordered Statistics postprocessing step Decoder (BP-OSD) as introduced in Refs.~\cite{roffe2020decoding,panteleev2021degenerate} for physical error rates $p$ down to $10^{-3}$ and then extrapolate for lower error rates following the fitting formula of the main text.

\begin{figure}[!h]
     \centering
     \includegraphics[width=0.33\textwidth]{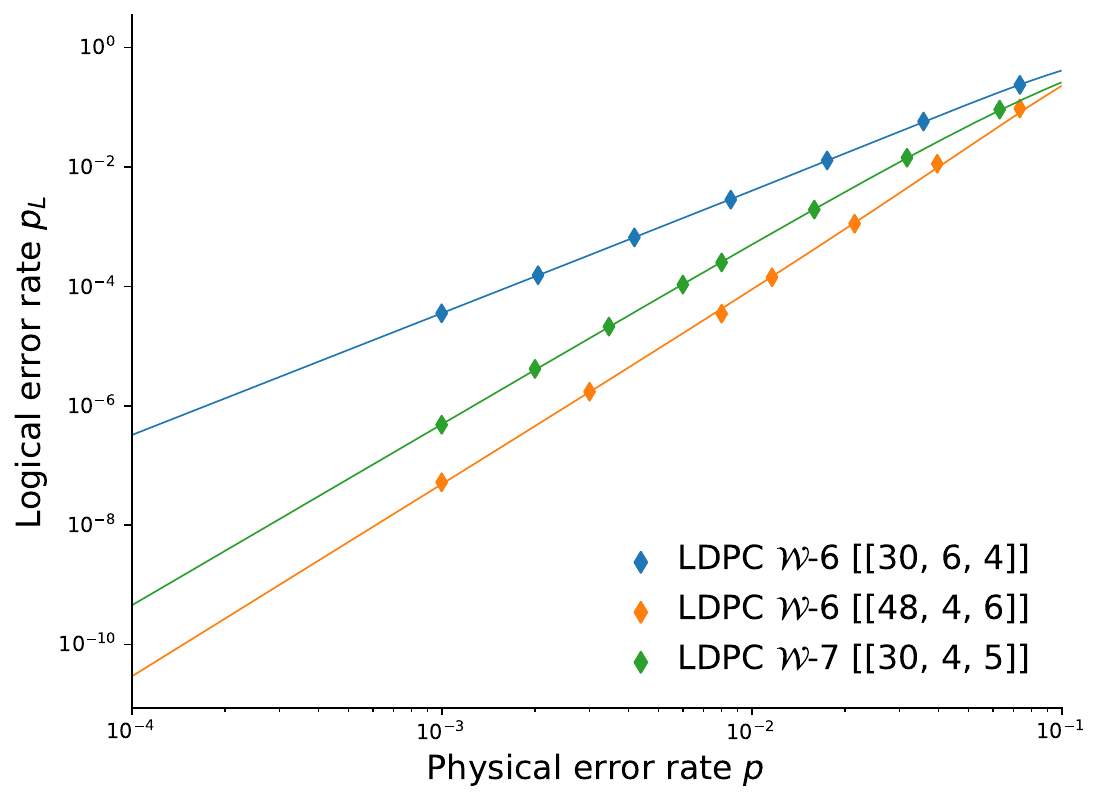}
     \caption{Comparison of noise properties between TB codes of of weight-$6$ and weight-$7$.
     We find the fitted code distances $d_\text{fit}=\{4.1, 6.4, 6.1\}$, respectively. 
     }
     \label{fig:poster_codes_weight67}
 \end{figure}

In Fig.~\ref{fig:poster_codes_weight67}, study codes of stabilizer weights six and seven.

\begin{figure}[!h]
    \centering
 \includegraphics[width=0.4\textwidth]{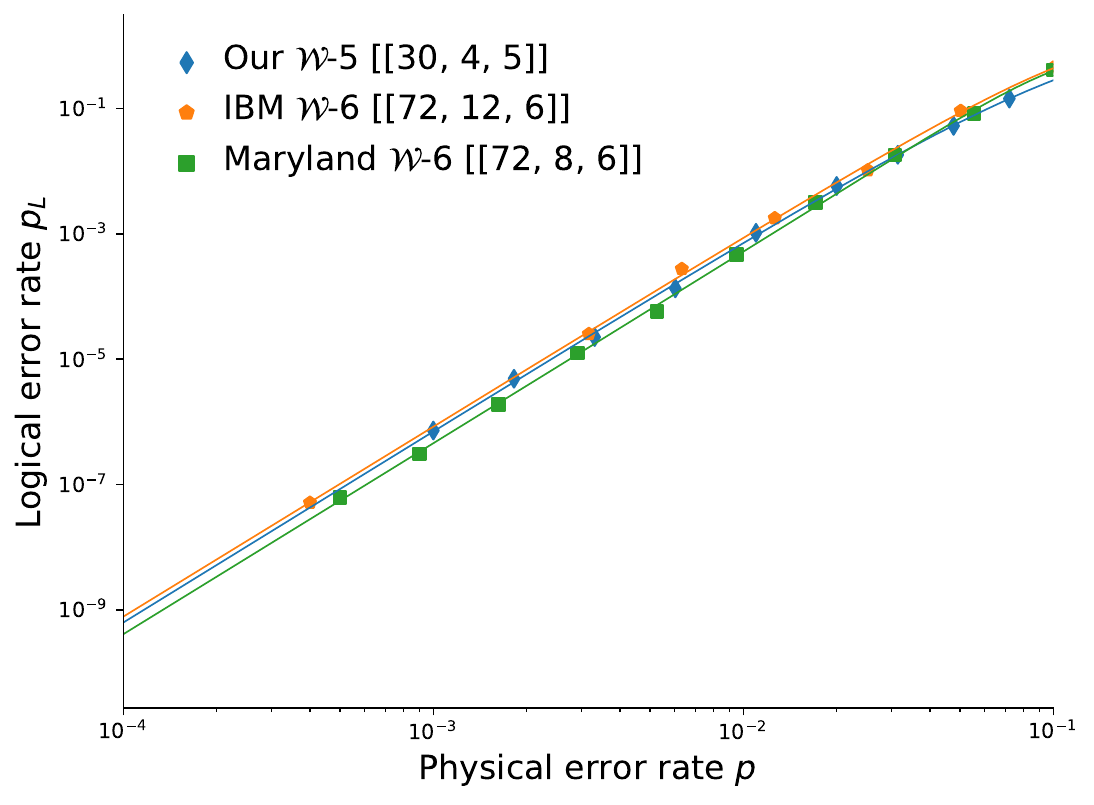}
    \caption{Measurement-noise free logical error rate $p_\text{L}$ of our weight-5 $\dsl 30, 4, 5\dsr$ TB code, weight-6 $\dsl 72,8,6\dsr$ BB code of Ref.~\cite{Berthusen_2024}, and weight-6 $\dsl 72,12,6\dsr$ BB code of Ref.~\cite{Bravyi_2024}.} 
\label{fig:comparison_IBM_Maryland_d6}
\end{figure}

In Fig.~\ref{fig:comparison_IBM_Maryland_d6}, we compare the noise suppression characteristics of our weight-$5$ $\dsl 30,4,5\dsr$ TB code with the weight-6 $\dsl 72,12,6\dsr$ code of Ref.~\cite{Bravyi_2024} and the weight-6 $\dsl 72,8,6\dsr$ code of Ref.~\cite{Berthusen_2024}. Here, we assume that the measurement of the syndrome is noise-free.

\section{Selected Codes of different weight}
\begin{figure}[!h]
    \begin{minipage}[b]{0.48\linewidth}
        \includegraphics[width=\linewidth]{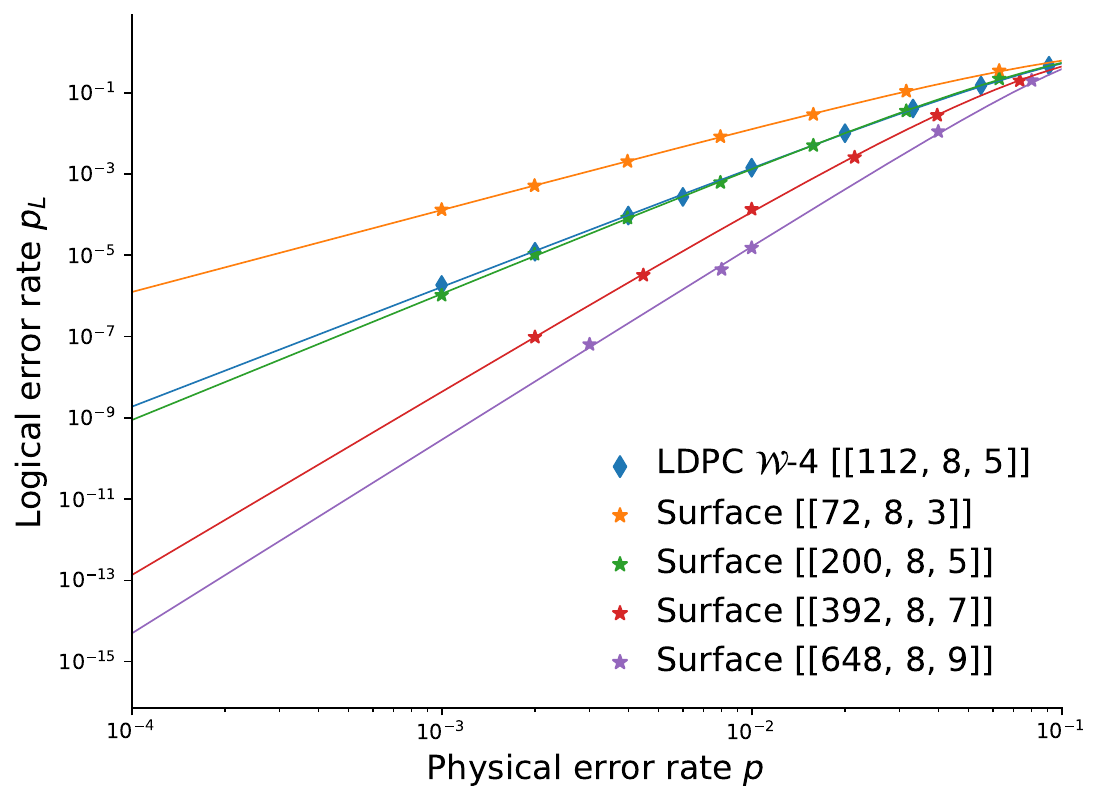}
        \caption{Logical error rate $p_\text{L}$ against physical error rate $p$ for our weight-4 code $\dsl 112, 8, 5\dsr$, and surface codes with similar code parameters. The error curves are fitted with Eq. (4) of the main text, yielding a fitted code distance of $d_\text{fit}=5.9$.}
        \label{fig:weight4_112-8-5}
    \end{minipage}
    \hfill %
    \begin{minipage}[b]{0.48\linewidth}
        \includegraphics[width=\linewidth]{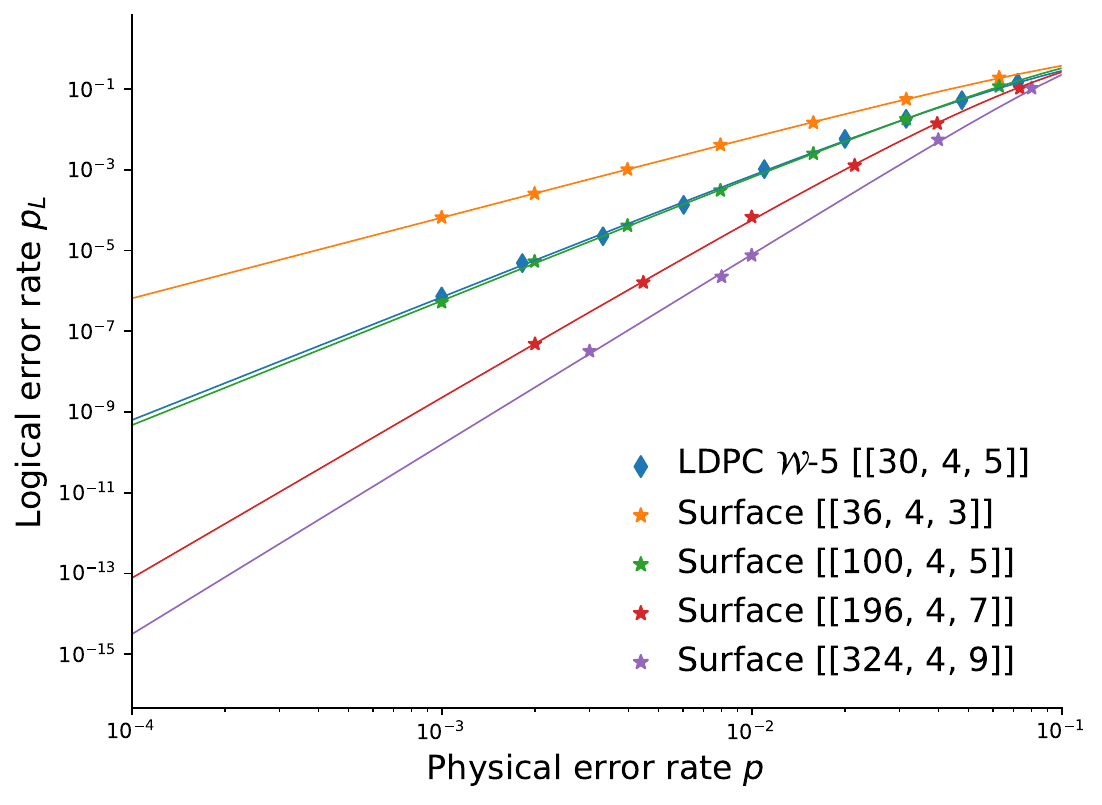}
        \caption{Logical error rate $p_\text{L}$ against physical error rate $p$ for our weight-5 code $\dsl 30, 4, 5\dsr$. We show surface codes with comparable code parameters as reference. We find $d_\text{fit}=6.1$ for our QLDPC code.} 
        \label{fig:decoding_weight5_30-4-5}
    \end{minipage}
\end{figure}

\prlsection{Weight-4}
In Fig.~\ref{fig:weight4_112-8-5}, we show the physical error rate $p$ against the logical error rate $p_\text{L}$ for our weight-4 QLDPC code $\dsl 112, 8, 5\dsr$. As comparison, we also show surface codes with the same number of logical qubits and similar distance $d$.  We find that our code has comparable error suppression than the $\dsl 200,8,5\dsr$ surface code, while requiring only about half the number of physical qubits. We also fit the logical error rates with Eq.\,(4) in the main text, where for our code with find a fitted code distance $d_\text{fit}=5.9$. 

\prlsection{Weight-5} 
Next, in Fig.~\ref{fig:decoding_weight5_30-4-5} we study the error suppression of our $\dsl 30, 4, 5\dsr$ code, which is a weight-$5$ TB code.
Our code suppresses logical errors at a physical error rate $p=10^{-3}$ by more than three orders of magnitude, allows for a toric bi-planar layout and has about three times higher encoding rate compared to the $\dsl 100,4,5\dsr$ surface code which has similar error suppression.
\begin{figure}[!h]
    \begin{minipage}[t]{0.48\linewidth}
        \includegraphics[width=\linewidth]{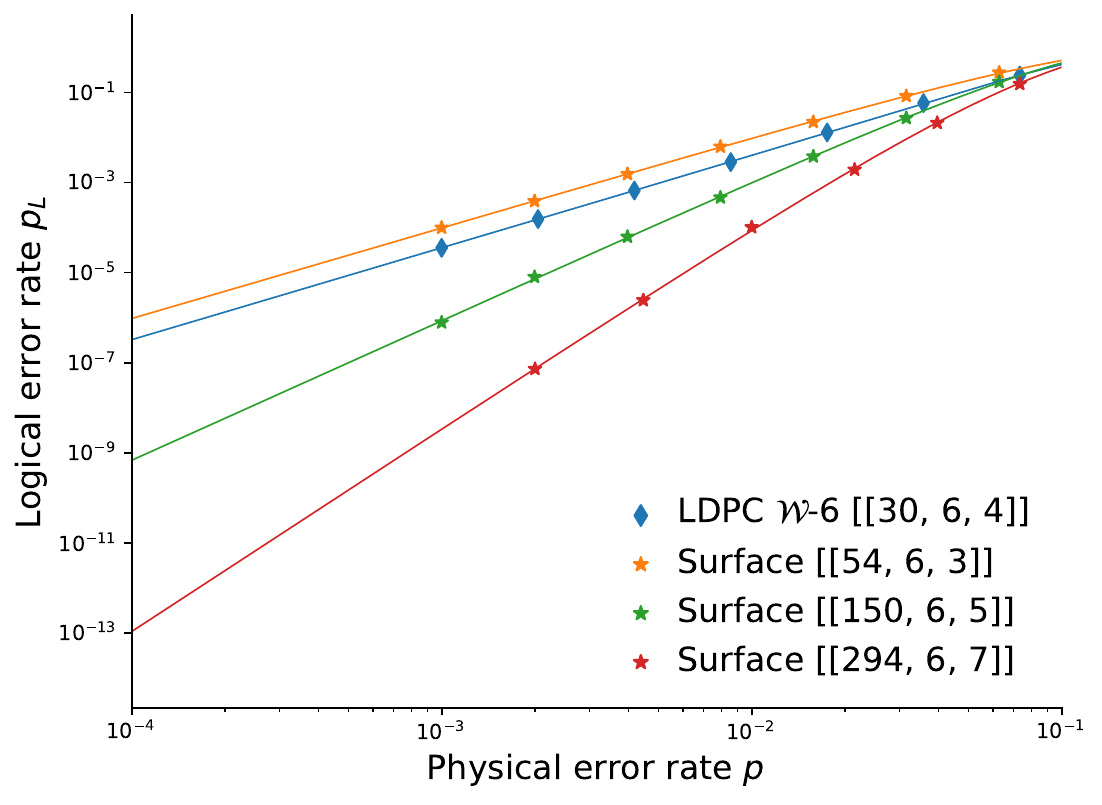}
        \caption{Logical error rate of our weight-6 code $\dsl 30, 6, 4\dsr$ against comparable surface codes. We find a fitted code distance of $d_\text{fit}=4.1$.}
        \label{fig:decoding_plot_weight6_30-6-4}
    \end{minipage}
    \hfill %
    \begin{minipage}[t]{0.48\linewidth}
        \includegraphics[width=\linewidth]{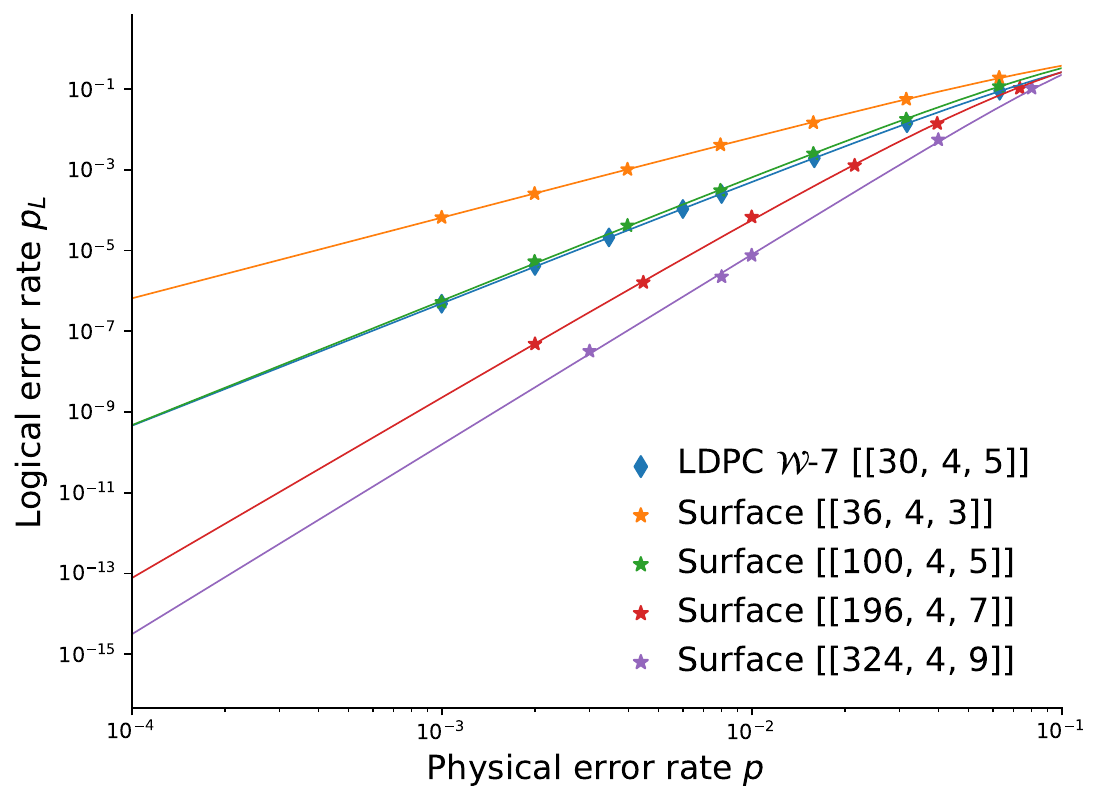}
        \caption{Our weight-7 code $\dsl 30, 4, 5\dsr$ compared against surface codes with the same number of logical qubits. We find a fitted code distance of $d_\text{fit}=6.1$.}
        \label{fig:weight7_30-4-5}
    \end{minipage}
\end{figure}\\
\prlsection{Weight-6}
In Fig.~\ref{fig:decoding_plot_weight6_30-6-4}, we present our weight-6 $\dsl 30, 6, 4\dsr$ code which has a high encoding rate of $1/5$ and find it to have a toric layout with bi-planar structure. We find comparable error suppression scaling as the distance $3$ surface code.
At physical error rate $p=10^{-3}$, we find a logical error rate of $3.5 \times 10^{-5}$ corresponding to a logical noise suppression by two orders of magnitude. We note that a surface code with distance-$4$ and the same number of logical qubits requires $96$ physical qubits, making our code three times more qubit efficient.

\prlsection{Weight 7}
Now, in Fig.~\ref{fig:weight7_30-4-5}, we study our weight-7 code with parameters $\dsl 30, 4, 5\dsr$. Note that we also found a weight-$5$ code with the the same code parameters. We find that our weight-7 $\dsl 30, 4, 5\dsr$ code has similar error suppression than the $\dsl 100, 4, 5\dsr$ surface code.

\section{Decoding performance}\label{app::decoding plots}
\begin{figure*}[!htbp]
    \centering
    \begin{minipage}[b]{0.48\linewidth}
        \includegraphics[width=\linewidth]{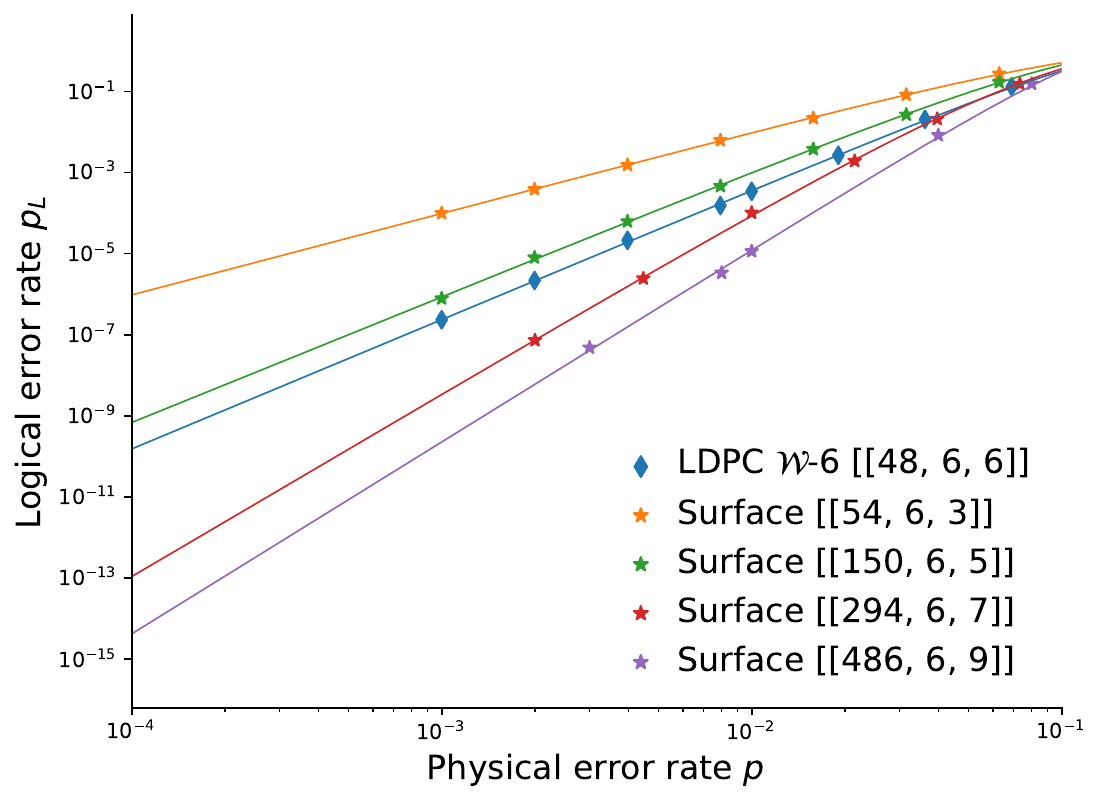}
        \caption{Our weight-6 code $\dsl 48, 6, 6\dsr$ has a lower logical error rate than the distance-$5$ surface code while requiring three times less qubits coming with an encoding rate of $1/8$.} 
        \label{fig:decoding_weight6_48-6-6}
    \end{minipage}
    \hfill %
    \begin{minipage}[b]{0.48\linewidth}
        \includegraphics[width=\linewidth]{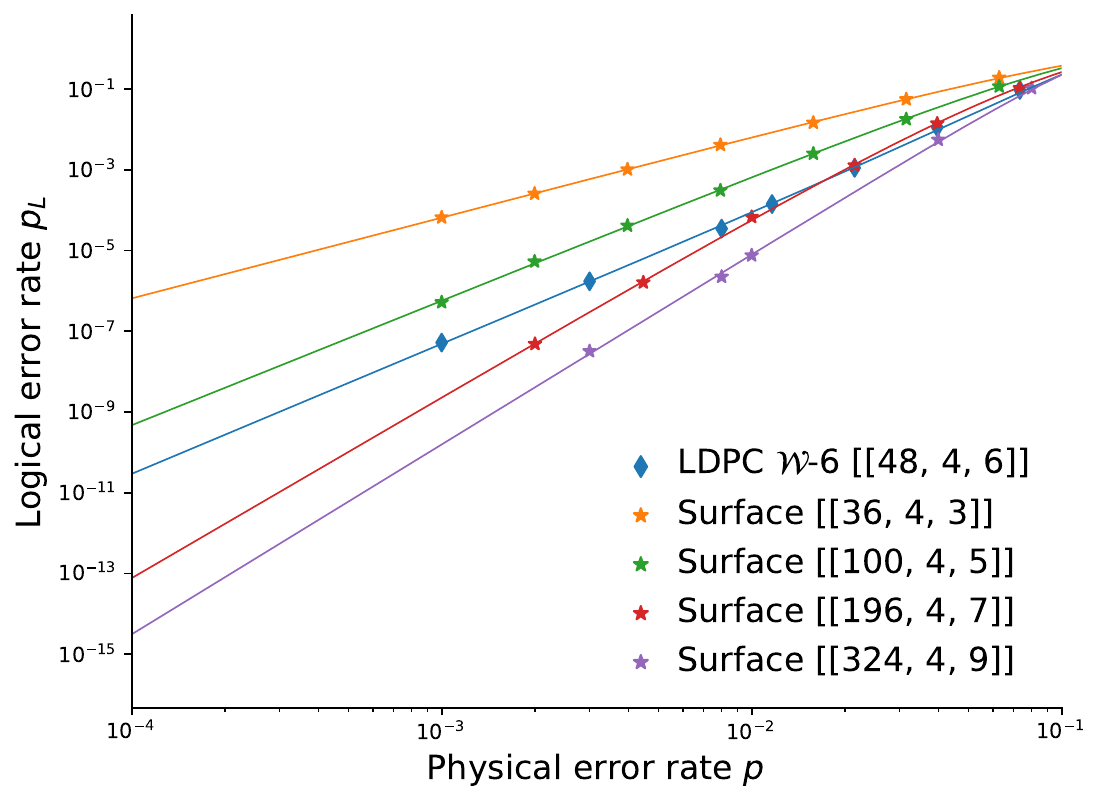}
        \caption{Our weight-6 code $\dsl 48, 4, 6\dsr$ with encoding rate of 1/12 only needs a third of the number of physical qubits compared to the distance-$6$ surface.}
        \label{fig:decoding_plot_weight6_48-4-6}
    \end{minipage}

    \begin{minipage}[b]{0.48\linewidth}
        \includegraphics[width=\linewidth]{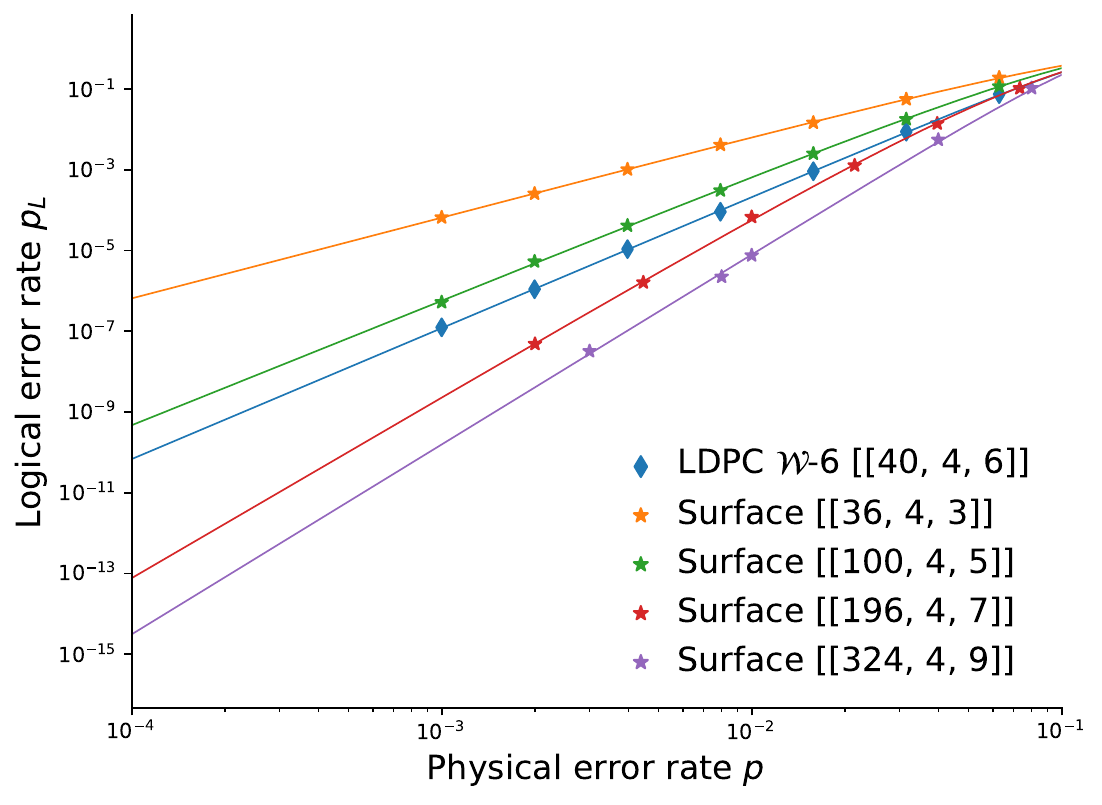}
        \caption{Weight-6 code with high encoding rate of $10\%$ and parameters $\dsl 40, 4, 6\dsr$. Compared to the distance-6 surface code, our code requires less physical qubits by a factor of $3.6$.}
        \label{fig:decoding_plot_weight6_40-4-6}
    \end{minipage}
    \hfill %
    \begin{minipage}[b]{0.48\linewidth}
        \includegraphics[width=\linewidth]{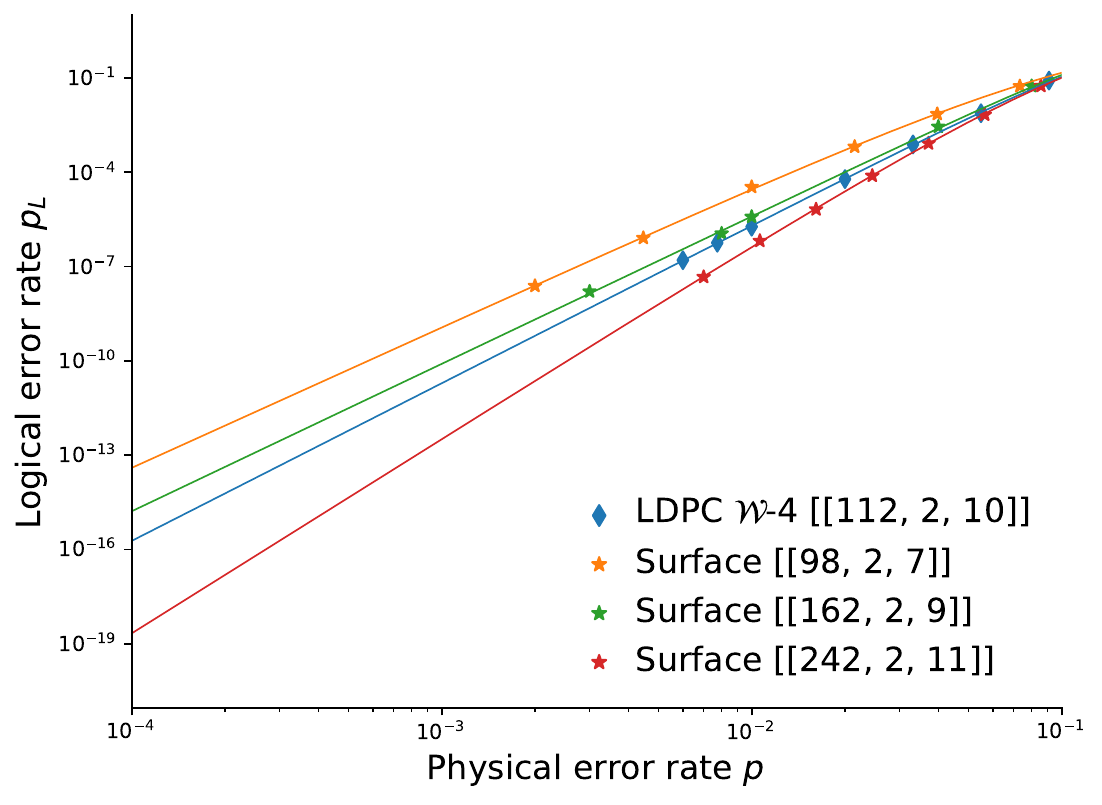}
        \caption{This weight-4 code $\dsl 112, 2, 10\dsr$ requires only about half the number of physical qubits compared to the surface code at similar noise suppression performance.}
        \label{fig:decoding_plot_weight4_112-2-10}
    \end{minipage}
\end{figure*}

We show noise suppression performance for our QLDPC codes of weight six under depolarizing noise. In Figs.~\ref{fig:decoding_weight6_48-6-6} -~\ref{fig:decoding_plot_weight4_112-2-10}, we show our codes in comparison with a selection of surface code of lower and higher distance than the QLDPC code to illustrate the noise performance. Since a surface code architecture would require $k$ batches of a $d\!\times\!d$ grid of qubits, the total number of physical qubits required to implement $k$ logical qubits at code distance $d$ reads $n = kd^{2}$.

\begin{figure*}[!htbp]
    \centering
    \begin{minipage}[b]{0.48\linewidth}
        \includegraphics[width=\linewidth]{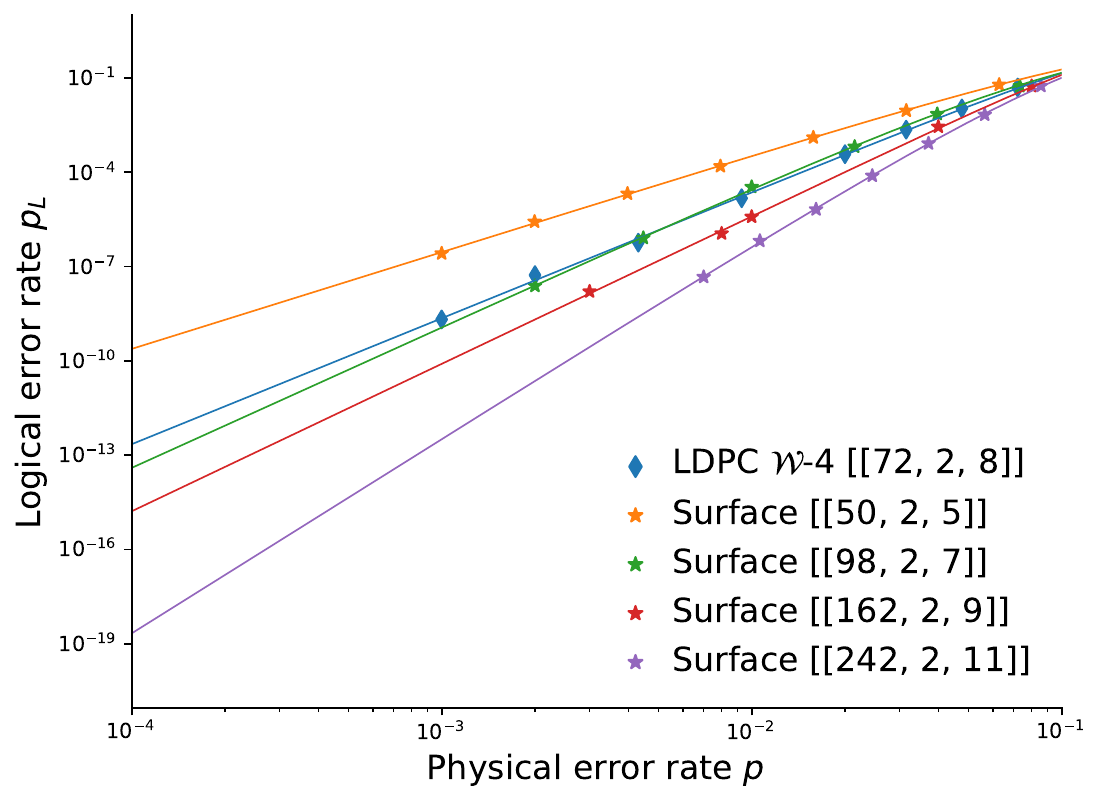}
        \caption{Our weight-4 code $\dsl 72, 2, 8\dsr$ offers a high code distance while just using almost half of the number of physical qubits than the surface code would require.}
        \label{fig:decoding_plot_weight4_72-2-8}
    \end{minipage}
    \hfill %
    \begin{minipage}[b]{0.48\linewidth}
        \includegraphics[width=\linewidth]{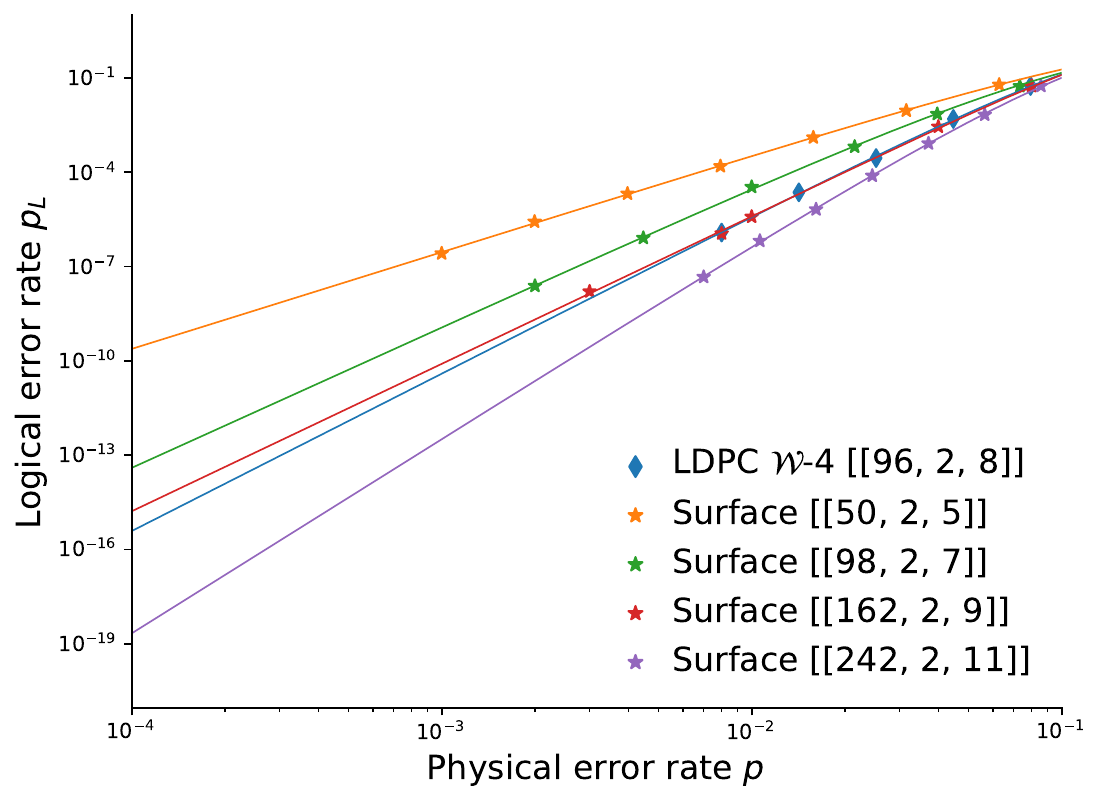}
        \caption{Our weight-4 code $\dsl 96, 2, 8\dsr$ provides good noise suppression characteristics and requires less physical qubits than the equivalent surface by $33\%$.}
        \label{fig:decoding_plot_weight4_96-2-8}
    \end{minipage}

    \begin{minipage}[b]{0.48\linewidth}
        \includegraphics[width=\linewidth]{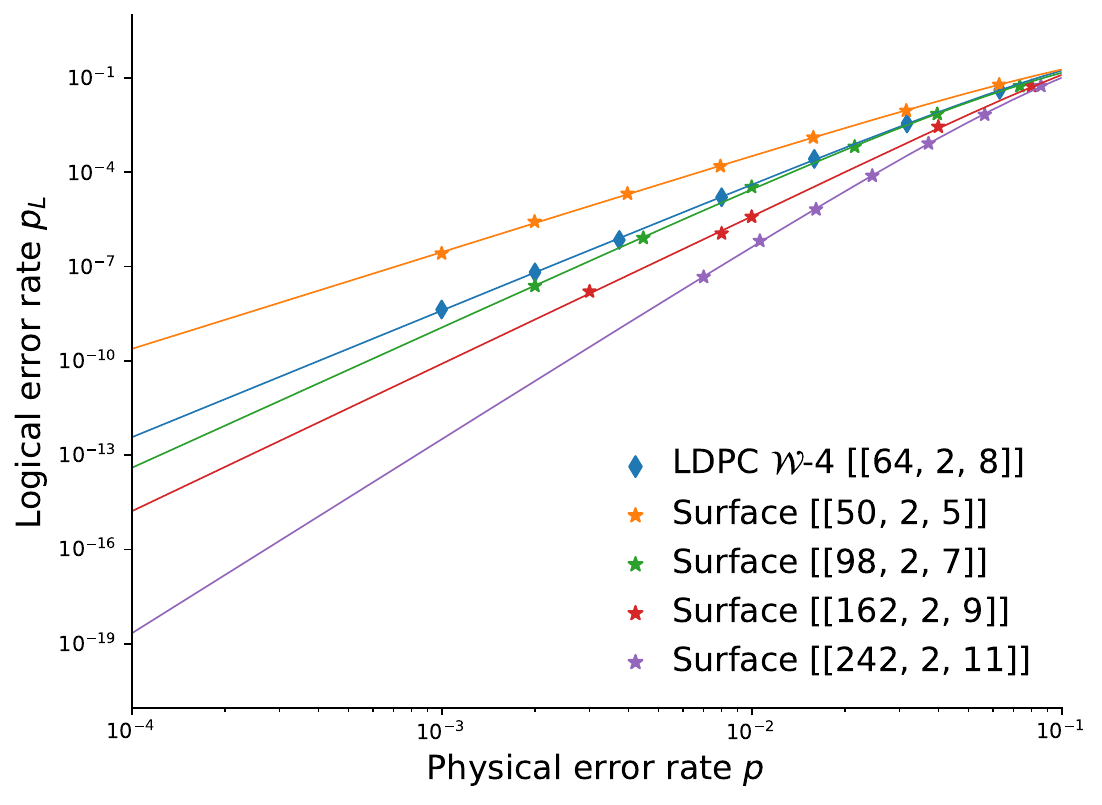}
        \caption{Our weight-4 code $\dsl 64, 2, 8\dsr$ encodes two logical qubits at only half the resources needed for a surface code implementation.}
        \label{fig:decoding_plot_weight4_64-2-8}
    \end{minipage}
\end{figure*}

\newpage
Next, we show the noise suppression performance for QLDPC codes of weight four under depolarizing noise. In Figs.~\ref{fig:decoding_plot_weight4_72-2-8} -~\ref{fig:decoding_plot_weight4_64-2-8}, we show our codes in comparison with a selection of surface code of lower and higher distance than the QLDPC code to illustrate the noise performance. Since a surface code architecture would require $k$ batches of a $d\!\times\!d$ grid of qubits, the total number of physical qubits required to implement $k$ logical qubits at code distance $d$ reads $n = kd^{2}$.

\begin{figure*}[!htbp]
    \centering
    \begin{minipage}[b]{0.48\linewidth}
    \includegraphics[width=\linewidth]{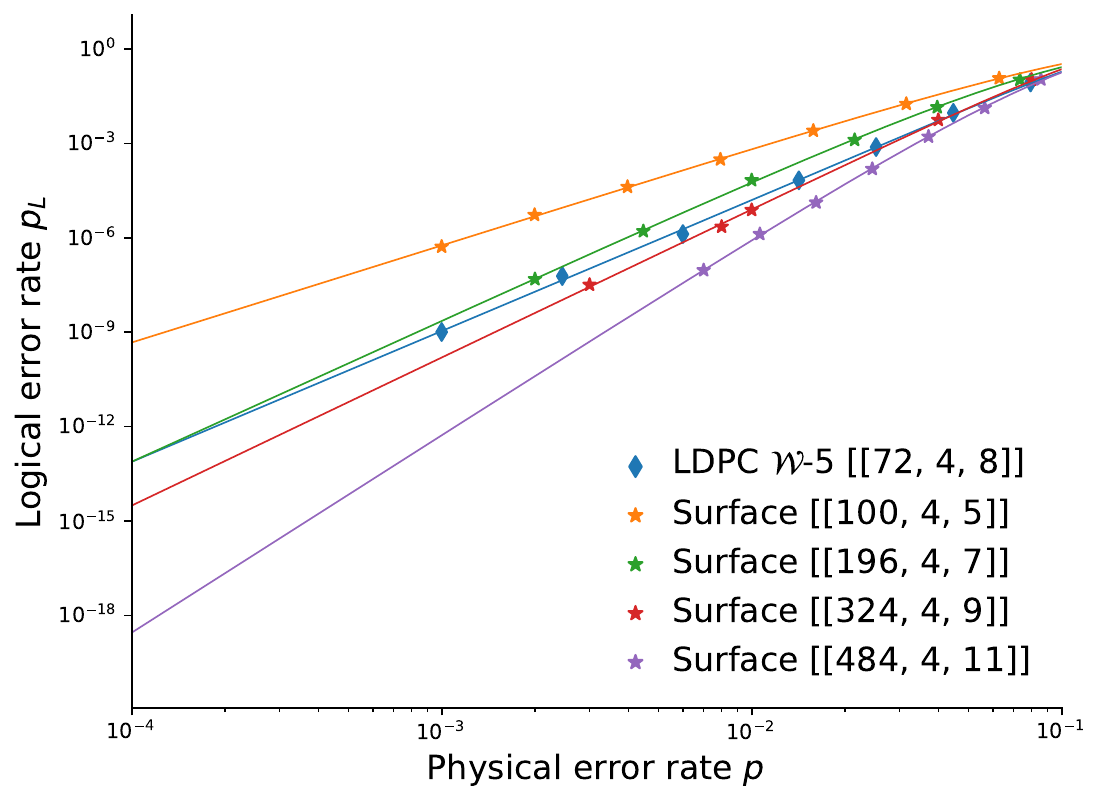}
            \caption{Our weight-5 code $\dsl 72, 4, 8\dsr$ shows better noise performance than the distance-$7$ surface code and is more qubit efficient by a factor of $2.7$. We find a fitted code distance of $8.3$.} 
        \label{fig:decoding_weight5_72-4-8}
    \end{minipage}
    \hfill %
    \begin{minipage}[b]{0.48\linewidth}
        \includegraphics[width=\linewidth]{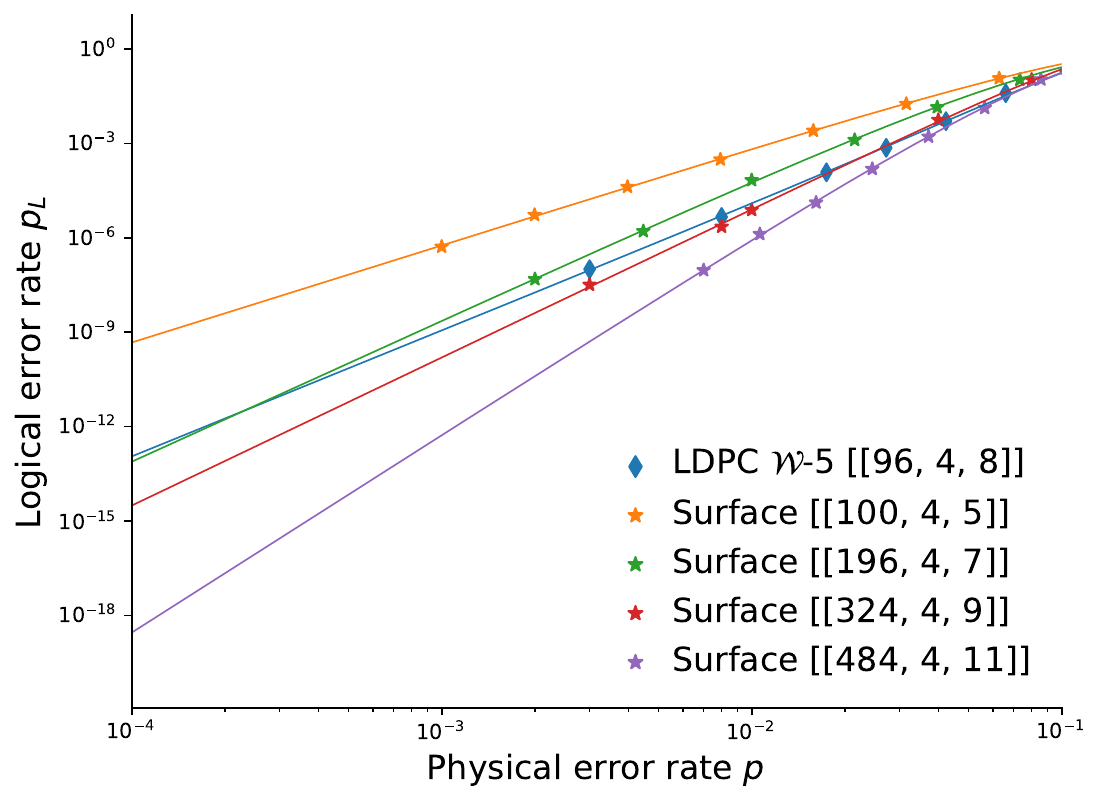}
        \caption{With our weight-5 code $\dsl 96, 4, 8\dsr$ we provide a mix out of high code distance and moderate encoding rate. Further, an implementation requires less physical qubits than the distance-8 surface code by almost a factor of $2.7$.}
        \label{fig:decoding_plot_weight5_96-4-8}
    \end{minipage}
\end{figure*}

\newpage
Finally, we show the suppression performance for QLDPC codes of weight five under depolarizing noise. In Figs.~\ref{fig:decoding_weight5_72-4-8} -~\ref{fig:decoding_plot_weight5_96-4-8}, we show our code in comparison with a selection of surface code of lower and higher distance than the code to illustrate the noise performance. Since a surface code architecture would require $k$ batches of a $d\!\times\!d$ grid of qubits, the total number of physical qubits required to implement $k$ logical qubits at code distance $d$ reads $n = kd^{2}$.

\clearpage
\newpage

\section{Tanner Graph Details}\label{app::tannergraph} %

Here, we recall the definitions used in the main text for reader convenience, but further details can be found in a standard reference such as~\cite{Bondy2008}.
We use the terms `physical' or `data' qubits to interchangeably.

Before diving into any technical details, it is important to point out that the physical implementation of a QECC does not have to directly conform to the simplest Tanner Graph structure, as along as the graph connectivity is retained. One can arbitrarily permute the qubits around physical space as they wish, changing the physical implementation of the code and the corresponding stabilizers' supports. However, a symmetric Tanner Graph structure can greatly ease hardware implementation, by noting that the Tanner Graph embedded in D-dimensional Euclidean Space $\mathbb{R}^D$ directly suggests one choice of physical implementation, formalized by Proposition \ref{proposition:TannerPhysical}. Therefore, we seek the simplest possible embedding in physical space, which in the case of MB codes is the toric layout.

Recall that the Tanner Graph of any CSS code is a bipartite graph $(V_{check} \cup V_{data}, E)$, such that any edge e connects some physical qubit $v_{data}$ to some check $v_{check}$. Bipartite here refers to the fact that there are no connections between check vertices, and similarly no checks between physical qubit vertices. Checks and their action on physical qubits are completely specified by the Tanner Graph, thus specifying the Tanner Graph specifies a code.

Given a graph G, we now rewrite vertices of G into indices $\{i\}$, and the edges of G into a matrix, called the adjacency matrix A, where $a_{ij}=1$ when there is an edge between vertex i and vertex j, 0 otherwise. Thus, the $A_i$ and $B_j$ terms that make up A and B in our definition of a MB-QLDPC code, give rise to the adjacency matrix of the code's Tanner Graph.

We formally define an embedding as a map E: G $\rightarrow$ $(\mathbb{R}^D)^{|V|} \times (\mathbb{R}^D)^{|E|}$. The first argument of $(\mathbb{R}^D)^{|V|} \times (\mathbb{R}^D)^{|E|}$ will specify the locations of the $|V|$ vertices as distinct points, and the second argument will specify the location of the $|E|$ edges as continuous curves.

An interaction vector (i, j) of a Tanner Graph embedding is defined as a vector that starts from data qubit vertex and ends on a check qubit vertex, the two being connected by an edge.

Define a qubit two-block CSS code as CSS codes such that $H_X = [A|B]$ and $H_Z = [B^T |A^T ]$, where A and B commute and have the same size. Two-blocks allow us to partition data qubits into L and R qubits. TB codes are a special type of two-block CSS code, amongst others that have been studied where A and B arise from a group algebra \cite{lin2023quantum}. 

Proposition~\ref{proposition:TannerPhysical} below should be well known amongst specialists in the field, but we state it for clarity.

\begin{proposition}[Tanner Graph Physical Implementation] 
\label{proposition:TannerPhysical}
Suppose $\exists M \in \mathbb{R}$, $\exists$ an embedding E: G $\rightarrow$ $(\mathbb{R}^D)^{|V|} \times (\mathbb{R}^D)^{|E|}$ of the Tanner graph with an upper bounded vertex density $\rho \leq M$ (vertices per unit volume) in $\mathbb{R}^D$, such that $\forall$check vertex $v_S$, every data qubit sharing an edge with $v_S$ is within distance $r_S$.

Then E provides a physical implementation of qubits of the code in $\mathbb{R}^D$ such that the support of each stabilizer S lies in within a ball of radius $r_S$, determined by the embedded Tanner Graph. 
\begin{proof}
The point of having an upper bounded vertex density $\rho$ is so that we don't try to `game' the system and place an increasing number of vertices per unit volume, which would allow a pathological example such as an arbitrarily large expander graph to embed into $\mathbb{R}^D$. In principle, we could allow for such cases but they are experimentally impractical above some density $\rho$ due to squeezing too many qubits near one another. Essentially, we would like to focus on a physically realistic tiling of qubits on $\mathbb{R}^D$.

We choose the physical embedding E of qubits provided in the hypothesis, such that each stabilizer S is supported in its ball of radius $r_S$, specifically on those data qubit vertices that share an edge with the check vertex of S. In our paper, this refers to the toric layout embedded on a lattice in $S^1 \times S^1$, and indeed all for our weight-5 $\dsl 30,4,5\dsr$ code in Fig 1 of the main text, our stabilizers will be supported on 4 neighbouring qubit sites, and 1 long-range qubit site.
\end{proof}
\end{proposition}

Note that the above proposition also implies that for any QECC such that its Tanner Graph has a local embedding in $\mathbb{R}^D$, the QECC parameters $\dsl n,k,d\dsr$ must satisfy the Bravyi-Poulin-Terhal bound~\cite{Bravyi_2010}, the statement that for geometrically local codes, $\exists$ geometric locality constant $c$ such that $kd^\frac{2}{D-1} \leq cn$. This bound holds even if one chooses a non-local physical implementation. Thus, one should remain cautious about the performance of a code, even when provided with a non-local embedding of its Tanner Graph, as it may be graph isomorphic to a geometrically local code, where the notion of a graph isomorphism is defined below in Definition~\ref{definition:graphisomorphism}.

\begin{definition}[Planar]
Let \( G = (V, E) \) be a graph. \( G \) is said to be \textit{planar} if there exists an embedding E : G $\rightarrow (\mathbb{R}^2)^{|V|} \times (\mathbb{R}^2)^{|E|}$ such that:
\begin{enumerate}
    \item No curves intersect except at their endpoints.
    \item No point is included in more than one arc except as an endpoint.
\end{enumerate}
\end{definition}

\begin{definition}[Subgraph]
A subgraph S = $(V_S, E_S)$ of G = $(V,E)$ is a graph such that $V_S \subset V$ and $E_S \subset E$. In other words, every vertex and edge of S is a vertex and edge of G respectively.
\end{definition}

\begin{definition}[Thickness]
The thickness $\theta$ of a graph G = $(V,E)$ is the minimum number of planar subgraphs $\{S_i\}_{i = 1,...,\theta}$, where each $S_i = (V_i,E_i)$, such that the edges of G can be partitioned into such that $\cup_i V_i = V$ and $ \cup_i E_i = E$. 
\end{definition}

\begin{proposition}[Bi-planar Architecture] 
All TB-QLDPC codes of weight-4, where A = $A_1 + A_3$, B = $B_1 + B_3$, all codes of weight-5 such that A = $A_1 + A_3$, B = $B_1 + B_2 + B_3$, and all codes of weight-6 such that  A = $A_1 + A_2 + A_3$, B = $B_1 + B_2 + B_3$ (as presented in ~\cite{Bravyi_2024}), or A = $A_1 + A_3$, B = $B_1 + B_2 + B_3 + B_4$ allow for a bi-planar architecture of thickness $\theta = 2$, with a bi-planar decomposition attained in time $O(n)$.
\label{proof:double_toric_structure_weight4}
\begin{proof}

For cases weight-4 and 5, we can directly carry over `planar-wheel' proofs for the degree-6 graphs in Lemma 2 of~\cite{Bravyi_2024}, where in their work A = $A_1 + A_2 + A_3$ and B = $B_1 + B_2 + B_3$, by setting $A_2$ = $B_2$ = 0 for weight-4 codes, and setting just $A_2$ = 0 for weight-5 codes. We will also have an identical proof structure, just with empty edges on locations where the terms are set to zero. 

For cases weight-6 where A = $A_1 + A_3$, B = $B_1 + B_2 + B_3 + B_4$, we now partition G into two subgraphs $G_\gamma$ and $G_\eta$ where the CSS codes representing subcodes $\gamma$ and $\eta$ have parity check matrices as follows:
\begin{equation}
    H^X_{\gamma} = [A_1|B_1+B_2] \quad \text{and} \quad H^Z_{\gamma} = [B_1^T+B_2^T|A_1^T] \quad \text{,} \quad H^X_{\eta} = [A_3|B_3+B_4] \quad \text{and} \quad H^Z_{\eta} = [B_3^T+B_4^T|A_3^T]
\end{equation}
This allows a bi-partition into two degree-3 subgraphs, allowing us to adapt the proof of~\cite{Bravyi_2024} on each degree-3 subgraph of our weight-6 codes.
\end{proof}
\end{proposition}

Note that the above proof breaks down if we have $A = A_1 + A_3, B = B_1 + B_2 + B_3 + B_4 + B_5$ such as in our weight-7 code. It is easy to extend the above argument to show that weight-7,8,9 codes have a thickness-3 implementation (and analogously for larger weights and thickness), by splitting into 3 CSS subcodes, but high thickness is not desirable for near-term hardware implementation. The Tanner Graph is degree-7, but it is known that degree-7 thickness-2 graphs exist, such as the complete $K_7$ graph, defined as having 7 vertices and all-to-all connectivity~\cite{Bondy2008}. Furthermore, a lower bound on graph thickness is obtained from a corollary of Euler's polyhedron formula, that given a graph with $p$ vertices and $q$ edges, $\theta \geq \frac{q}{3p-6}$~\cite{Mutzel1998}, yielding a lower bound of $\theta \geq \frac{25}{29}$ for our weight-7 $\dsl 30, 4, 5\dsr$ codes. It is easy to verify that the graph is not planar, but we could remain hopeful that it has thickness $\theta = 2$.

In order to rule out the possibility that the graph may fortunately be of thickness $\theta \!=\! 2$, a numerical program is necessary. Furthermore, by explicitly checking subgraphs, we will obtain an bi-planar layout for experimental realization. This brute-force procedure for checking $\theta \!=\! 2$ is generically NP-complete~\cite{Mansfield_1983}. For very small codes such as the Shor Code~\cite{Shor_1995}, it is still practical to enumerate decompositions into 2 subgraphs, and then check the planarity of each subgraph. However, our weight-7 $\dsl 30, 4, 5\dsr$ code did not provide an answer after a few days of simulation. An upper bound provided below on the computational complexity of a brute-force search for bi-planarity suggests that the difficulty of iterating over subgraphs is the key area for improvement in aiming towards a $\text{poly}(n)$ time complexity. 

\begin{proposition}[weight-$\mathcal{W}$ QLDPC Graph Bi-planarity Test Complexity]
\label{proposition:bi-planarcomplexity}
For any QLDPC code of sparsity $\mathcal{W}$, recall that its Tanner Graph will have degree $\mathcal{W}$. Then testing for bi-planarity is $\mathcal{O}(n \, 2^n)$.

\begin{proof}
We analyse the main steps in testing of bi-planarity:
\begin{enumerate}
    \item Breaking the Tanner Graph into 2 subgraphs allows for $\sum_{i = 1,...,n}$$ n \choose i $ = $2^{n}$ choices of subgraphs.
    \item Looping over each subgraph choice, we test for planarity using the Left-Right Planarity Test with $\mathcal{O}(e)$ complexity~\cite{DE_FRAYSSEIX_2006}, where $e$ is the number of edges of the graph being tested for planarity. In the Tanner Graph with sparsity $\mathcal{W}$, we have $e = \mathcal{O}(\mathcal{W}n)$, thus the planarity test for both subgraphs is $\mathcal{O}(n)$.
\end{enumerate}

Putting them together, we get an upper bound of $\mathcal{O}(n 2^n)$, an exponential time complexity.
\end{proof}
\end{proposition}

At this point, it is worth noting that we could generically define a $b$-block group algebra code as $[A_1 | A_2|...|A_b]$, enforce that $\sum_i A_i A_{k-i+1}^T = 0$ to enforce the CSS condition, and study it as a qLDPC code. We expect the generic framework to carry over.

\begin{definition}[Graph Isomorphism]
\label{definition:graphisomorphism}
Two graphs \(G\) and \(H\) are isomorphic if there exists a bijection 
\[ g: V(G) \rightarrow V(H) \]
from the vertices of \(G\) to the vertices of \(H\)—such that any two vertices \(u\) and \(v\) of \(G\) have an edge in \(G\) $\iff$ \(g(u)\) and \(g(v)\) have an edge in \(H\). 

Intuitively, we can relabel the vertices and edges of G to get those of H. This is what allows us to view the Tanner Graph of MB-QLDPC codes on the discrete torus.
\end{definition}

Physical qubits are now bipartitioned into L and R qubits, $\frac{n}{2}$ = lm qubits each. L and R refer to the left and right blocks in the check matrix definition, $H_{X} = [A|B] \quad \text{and} \quad H_{Z} = [B^{T}|A^{T}]
$. 

A Cayley Graph is essentially the `natural graph structure' endowed onto any group, defined through the generators of a group. We focus on undirected Cayley Graphs, which have undirected edges.

As in Ref ~\cite{Bravyi_2024}, we re-define physical qubits and checks as each labelled by monomials, $M = \{x^i y^j \, \vert \, i = 0,1,.., l-1, j = 0,1,...,m-1\}$. We view the monomials $M$ as a multiplicative group isomorphic (by a map $f^{-1}$ below) to the additive group of $\mathbb{Z}_{lm}$ = $\{0,1,...,lm-1\}$. Visually, we can represent both objects ($M$ and $\mathbb{Z}_{lm}$) by a planar x-y grid of l$\times$m points. These points on the grid are not qubits, but rather a collection of 4 objects, which consists of an L qubit, an R qubit, an X-check, a Z-check. Thus, this notation assigns each qubit or check a label q(T,$\alpha$), where T$\in \{L,R,X,Z\}$ is the data type and $\alpha \in M$ is the monomial label.

\[f : (+,\mathbb{Z}_{lm}) \rightarrow (\times,M), f(i) = x^{a_i} y^{i-ma_i} \quad \text{where} \quad a_i := \bigg\lfloor \frac{i}{m} \bigg\rfloor\] 
This isomorphism basically indexes along the x-axis from 0 to $l-1$ and along the y-axis from 0 to $m-1$. Multiplying by x corresponds to addition by m in $\mathbb{Z}_{lm}$, moving points along the x-y grid horizontally by a unit. Multiplying by y corresponds to addition by 1 in $\mathbb{Z}_{lm}$, moving points along the x-y grid vertically by a unit.  Since our third variable $z = S_l \otimes S_m$ = xy is contained in $M$, we have left $M$ unchanged under the introduction of variable $z$, allowing us to modify Lemma 3 and 4 of Ref~\cite{Bravyi_2024} to study the Tanner Graph of our codes.

In this framework, $\forall$ points $\alpha \in M$ , q(L,$\alpha$) data qubits have the set of X-check neighbours $L_X = \{ q(X, A_i^T) \alpha | i = 1,..., \mathcal{W}_A \}$ an Z-check neighbours $L_Z = \{ q(Z, B_i^T \alpha) | i = 1,..., \mathcal{W}_B \}$. Similarly, q(R,$\alpha$) data qubits have the set of X-check neighbours $R_X = \{ q(X,B_i^T) \alpha | i = 1,..., \mathcal{W}_B \}$ and Z-check neighbours $R_Z = \{ q(Z,A_i) \alpha | i = 1,..., \mathcal{W}_A \}$.

\begin{definition}[Cayley Graph]
Given a group C = $\langle c \rangle$ with generators c, the Cayley Graph $\Gamma(C, \{c\})$ is defined as the graph such that
\begin{enumerate}
    \item Vertices are the elements of $c \in C$
    \item $\forall g \in C$, $\forall \!$ generator $c$, there is an undirected edge between $g$ and $gc$. 
\end{enumerate}

For $\mathbb{Z}_a \times \mathbb{Z}_b$, its Cayley graph is the Cayley Graph generated by $\{(1,0), (0,1)\}$. Its Cayley Graph can geometrically be interpreted as a discrete torus with non-contractible cycles of length a and b, vertices being the vertex of the torus, edges being the grid lines of the torus. (1,0) and (0,1) would represent a unit shift along the x-direction and the y-direction respectively.
\end{definition}

\begin{definition}[Spanning Subgraph]
\label{definition:spanningsubgraph}
A spanning subgraph S of a graph G is a subgraph that contains all the vertices of G, but not necessarily all the edges.
\end{definition}

The next point to address is showing translational symmetry of edges on this toric layout that emerge from the X and Z check vertices. Once we establish translational symmetry, we can find the long-range edges by first noting that the toric layout provides a re-indexing map of the vertices into tuples on the square grid, f : V$\rightarrow \{(i,j)|i=1,..,2\mu \quad \text{and} \quad j=1,..,2\lambda\}$. Given a vertex at position $(i, j)$, after identifying the position $(i', j')$ of the long-range edge's adjacent vertex, we obtain the interaction vectors ($i'-i$, $j'-j$). Note that X and Z checks can have different interaction vectors.

Recall the formal definition of an embedding E: G $\rightarrow$ $(\mathbb{R}^D)^{|V|} \times (\mathbb{R}^D)^{|E|}$. In the proposition below, we say that a graph G is overlapped onto another graph H when embeddings $E_G$ and $E_H$ are chosen such that $E_G(G)$ = $E_H(H)$ in $(\mathbb{R}^D)^{|V|} \times (\mathbb{R}^D)^{|E|}$. Note that two graphs G, H can be overlapped $\iff$ G, H are graph isomorphic.

\begin{proposition}[TB-QLDPC Toric Layout Criterion for arbitrary weight, generalised from Ref.~\cite{Bravyi_2024}] 
\label{toriclayoutexplain}
    A TB-QLDPC code (\text{QC}($A$, $B$)) of weight $\mathcal{W}=\mathcal{W}_A+\mathcal{W}_B$ has a toric layout $\iff \exists i, j \in \{1,..,\mathcal{W}_A\}, g, h \in \{1,..,\mathcal{W}_B\}$ such that 
\begin{enumerate}
    \item \(\langle A_i A_j^T, B_g B_h^T \rangle = M\) and
    \item \(\operatorname{ord}(A_i A_j^T) \operatorname{ord}(B_g B_h^T) = lm\).
\end{enumerate}
\end{proposition}

To explain the idea behind Proposition~\ref{toriclayoutexplain}, which gives our criterion for a toric layout, split our data qubits into two blocks through the definition below. \[ H_{X} = [A|B] \quad \text{and} \quad H_{Z} = [B^{T}|A^{T}] \] L represents first half of the qubits (left side of block matrix), R represents the second half (right side of block matrix).

Our goal is to place the vertices of our Tanner Graph on the discrete torus indexed by $\mathbb{Z}_{2\mu} \times \mathbb{Z}_{2\lambda}$, but we allow for edge crossings. The intuition is that we need to look for a uniform choice of edge label to traverse between X checks (say X and X') in a cyclic manner, similarly for Z checks. Thus we have to choose an $A_j^T$ to go from X to L, and $A_i$ for L to X', similarly a $B_h^T$ and $B_g$ for R physical qubits and Z checks. Therefore, $A_i A_j^T, B_g B_h^T$ are the various monomials generated by our individual terms in the A and B matrices, representing 2-step crossings between X checks and Z checks respectively on the Tanner Graph. We fix a given X check vertex as 1, the identity, and multiply by $A_i A_j^T$ until we get back to 1. Therefore, ord($A_i A_j^T) = \mu$, where the order of the element determines how many 2-steps are made in a direction, starting from an X check before we return to the same X check. A similar argument follows for $\text{ord}(B_g B_h^T) = \lambda$ and Z checks. Thus, the Tanner Graph can be viewed on the torus $\mathbb{Z}_{2\mu} \times \mathbb{Z}_{2\lambda}$.

\begin{proposition}[TB-QLDPC Toric Layout Edge Translational Invariance]
\label{proposition:translationinvar}

For any TB-QLDPC code (with Tanner Graph $G$) of sparsity $\mathcal{W}$ with a toric layout, for any vertex $v_T$ of a fixed type $T \in \{L, R, X, Z\}$, all of the $\mathcal{W}$ edges of $v_T$, including the long-range edges, have translationally invariant interaction vectors on the toric layout.

\begin{proof}
A concrete example of edge translational invariance can be seen in the example of Fig 1 in the main text. We first define two concepts relevant in our TB codes.

Circulant matrices $C$ are matrices such that if $(c_0,c_1,...,c_i)$ is row-r of the matrix, then $(c_i, c_0, c_1,...,c_{i-1})$ is row-$(r+1)$ of the matrix, the rows are shifted right by one position. Because a circulant matrix C can be diagonalized into C = FD$F^{-1}$ using a Discrete Fourier Transform F, where $\omega = e^{\frac{2\pi i}{N}}$ and the action of F is defined as F $\ket{j}$ = $\sum_{k = 1}^{N-1}  \omega^{-jk} \ket{k}$, we conclude that its k-th power $C^k = F^{-1} D^kF$ is also a circulant matrix.

Cyclic matrices are matrices such that if $\exists$ $(c_0,c_1,...,c_i)$ is a row of the matrix, then $(c_i, c_0, c_1,...,c_{i-1})$ is also a row of the matrix. Note that if A and B are circulant, then A$\otimes$B is cyclic.

We use the isomorphism provided above, recalled here for convenience. \[f : (+,\mathbb{Z}_{lm}) \rightarrow (\times,M), f(i) = x^{a_i} y^{i-ma_i} \quad \text{where} \quad a_i := \biggl\lfloor \frac{i}{m} \biggr\rfloor\]

On the 2-D x-y grid defined by the isomorphism f, as mentioned we view L and R data qubits, X and Z checks to tile the grid, 4lm vertices in total, with one combined $(L,R,X,Z)$ unit per point. We do not define edges between data qubits and check vertices yet.

Recall that the TB codes are defined using shift matrices x, y, z = $S_l \otimes I_m$, $I_l \otimes S_m$, $S_l \otimes S_m$. $S_l$ and $S_m$ are circulant matrices. Thus x, y, z are cyclic matrices.

Furthermore, $A_i$ being a power of some variable x, y, z, is a tensor product of circulant matrices, thus cyclic. Since $A_i$ is cyclic, for a fixed row r, it has non-vanishing entries on a set $\{(a_{i_1},...,a_{i_{W_A}})\}$, which is translationally-invariant on the x-y grid because $A_i$ is cyclic, thus $\exists$ r' such that $\{(a_{i_{W_A}}, a_{i_1},..., a_{i_{W_A-1}})\}$.

For each $A_i$ term introduced, we have for each vertex v an edge $e_{v,X}$ along the grid to some X-check, call it X. Similarly we have $e_{v,Z}$ to a Z-check called Z. Due to the $A_i$ being cyclic, the edges introduced are translationally invariant (with PBC) on the x-y grid defined by the isomorphism f. Note that at this point, we have not imposed any requirement of a toric layout structure.

We may add as many edges as we like by introducing more terms $A_i$ and $B_j$, and they remain translationally invariant by repeating the argument above.

Furthermore, now using our hypothesis that the code has a toric layout, we may now traverse along the edges that define the toric grid, as per Proposition 2 in the main tex to make them overlap on the Cayley Graph of $\mathbb{Z}_{2\mu} \times \mathbb{Z}_{2\lambda}$ embedded in $\mathbb{R}^2$ with PBC. Since the graph is translationally symmetric, all the edges have translationally symmetric interaction vectors, including the long-range edges.
\end{proof}
\end{proposition}

Such translation invariant codes generated by cyclic matrices have also been studied in the literature, such as bicycle codes and two-block quantum group algebra codes \cite{kovalev2013quantum, lin2023quantum}.

\begin{definition}[Twisted Toric Layout]
\label{definition:twisted_toric_layout}
A CSS QLDPC code C has a ($t_x$,$t_y$)-twisted toric layout $\iff \exists$ positive integers $t_x, t_y$ such that the Tanner Graph of C has a toric layout, upon imposing an additional shift of $t_x$ units along the x-direction of the periodic boundary conditions of the toric layout, similarly for $t_y$ along the y-direction. 
\end{definition}

Note the above definition has a positive or negative direction ambiguity. We fix it to be shifts along the positive direction.

By this definition, the codes with a toric layout such as in Fig 1 of the main text, will have $t_x$ = $t_y$ = 0, the trivial twist. Non-trivial twisted toric layouts no longer have global translational symmetry, but retain it in the bulk. Some work has been done on studying twisted layouts~\cite{sarkar2023graphbased}, and it can be fruitful to find a systematic way to incorporate such theoretical results in a search for desirable weight-4 QLDPC codes for near-term implementation, as a first step beyond the toric code.

\begin{definition}[Tangled Toric Layout]
\label{definition:tangled_toric_layout}
A CSS QLDPC code C has a ($\sigma$, $\tau$)-tangled toric layout $\iff \exists \sigma \in S_{2\mu}, \tau \in S_{2\lambda}$ where $S_q$ is the permutation group on $q$ elements, such that the Tanner Graph of C has a toric layout after applying $\sigma$ along the x-direction of the periodic boundary conditions of the toric layout, similarly for $\tau$ along the y-direction.
\end{definition}
The notion of tangled toric layout can be considered the most general case for disrupting the PBC of the torus, whilst keeping the horizontal and vertical PBC's separate. Note that the twisted layout definition is a special case of a tangled layout, corresponding to $\sigma$ being the permutation cycle that sends $i$ to $i+t_x$ mod $2 \mu$ $ \forall i$, and similarly $\tau$ sends $j$ to $j+t_{y}$ mod $2 \lambda$ $\forall j$.

We encourage readers to find systematic ways of studying tangling parameters $(\sigma, \tau)$ of codes that are surface codes in the bulk. As a first step in this direction, we provide one algorithm below for weight-4 TB codes that have a tangled toric layout.

\begin{proposition}[Algorithm for Weight-4 ($\sigma$, $\tau$)-Tangled Toric Layout Parameters]

For any two-block CSS QLDPC code of sparsity 4 with equal X and Z degree per data qubit (2 each), suppose it has a tangled toric layout known to have torus parameters $\mu$,$\lambda$.

Then there is an $O(n)$ time algorithm to find $\sigma$, $\tau$.
\label{proposition:two_block_tangled_layout}

\begin{proof}
It is instructive to refer to Fig 1 in the main text while reading this proposition, ignoring the long-range edges. The key idea is to embed a spanning subgraph onto the rectangular grid of size $2\mu \times 2\lambda$. The remaining edges on the boundary left unassigned to the square grid along x-direction and y-direction give us $\sigma$ and $\tau$ respectively.
We analyse the main steps:
\begin{enumerate}
\item A choice for the top-right vertex of the 2-D grid is made, it can be arbitrary. Fixing it as a data qubit, 2 distinct X and Z vertices are chosen to the left and below it. The other 2 are left unassigned, they are part of the boundary. 

\item Recall that by hypothesis, the CSS Tanner Graph is bipartite (into data qubits and checks) and degree-4. Our challenge now is to decide which data qubit type, L or R, should be placed along the horizontal axis of the embedding. A poor choice such as placing two L qubits diagonally adjacent to each other on the square grid, would prevent us from getting a layout like Fig 1 shown in the main text. This is where the L-R split of data qubits is crucial. Recall that we can bipartition L and R by using the left and right blocks in the check matrix definition, $H_X = [A|B]$ and $H_Z = [B^T |A^T ]$. The first half of the data qubit indices are in L, and the second half indices are in R. Thus, at an X check $v_X$, we choose the data qubit to its left to be an L qubit $v_L$, and the data qubit below it to be an R qubit, $v_R$. Finally, we place any Z-check ($v_Z$) that shares an edge with $v_R$ to the right of $v_R$. This allows us to produce a 2$\times$2 unit cell of the Tanner Graph embedding, consisting of 1 of each vertex type (X, Z, L, R).

\item We repeat the above assignment rules for each vertex by continuing to assign directions of data qubits and checks on the square grid based off the L-R split. As a result, the arrangement of all other vertices on the grid are uniquely specified (refer to Fig 1 of main text), by `snaking' our way from right to left, then moving to the next row, assigning vertices from the left to right, iteratively tessellating unit cells all the way down the 2-D grid. This step is $O(n)$ time complexity. 

\item By ignoring the bulk of the grid and only looking at the top and bottom edges of the 2-D grid defined along the x-direction boundary, we obtain a bipartite graph with two rows of vertices, and edges only between the rows, but not within each row. This bipartite graph is just a visual representation of a bijective map. This bijection defines a permutation map $\sigma$ for us upon indexing the qubits along the boundary by integers. There is no loop in this step of the algorithm, thus this step is additive in time complexity. This step is O($\sqrt{n}$) because we only enumerate the along the perimeter of the 2-D grid, which has O($\sqrt{n}$) vertices.

\item We repeat Step 4 along the y-direction for the left and right vertical edges, to obtain $\tau$.
\end{enumerate}

\end{proof}
\end{proposition}

The above algorithm also works for even more general notions of tangling, where the x and y boundary conditons may mix. In such cases, $(\sigma, \tau)$ parameters will be combined into a single tangling parameter $\gamma \in Bij(V_x \cup V_y)$, $\gamma$ being a permutation of the combined boundary vertices. $V_x$ is the vertices along the x-direction, similarly for $V_y$. 

The above proposition holds higher-weight codes, assuming additionally that one knows beforehand the choice of a spanning subgraph for a toric layout. However, this choice is generally not clear a priori. One would have to iterate through various choices of edges in the spanning subgraph, which is exponentially hard. For weight-4 codes, it suffices to know what $\mu$, $\lambda$ are, since the Tanner Graph is degree-4, leaving no freedom in the choice of edges on the spanning subgraph. Furthermore, it is easy to deduce $\mu$,$\lambda$ by using computer vision tools or visual inspection of the Tanner Graph, such as in Fig 4 shown in the main text.

One possible direction for further optimization of ultra low-weight codes is to explore the questions: What are the ideal forms of tangling for improving code parameters, decoding and hardware implementation? Concretely, this can be done by imposing further structure on tangling parameters $(\sigma, \tau)$ in a torus to improve code performance, of which a $(t_x, t_y)$ twisted toric layout is only one of many possibilities~\cite{sarkar2023graphbased}. We expect improved performance by a constant factor, but not in the asymptotic scaling.

\section{Syndrome measurement circuit}
We propose a depth-7 circuit to measure syndromes in weight 5 TB codes which is illustrated in Fig.1b in the main text. 
As this circuit interleaves X and Z checks, it is substantially shorter than a sequential measurement of X and Z. Numerically, we find that this circuit preserves the code distance even under circuit-level noise, in contrast to a simple sequential check.

However, one has to make sure that the interleaving of non-commuting checks leaves the overall parity check and logical operators invariant, which we show in the following, via the procedure outlined in Ref.~\cite{Bravyi_2024} for weight 6 BB codes.

Let us define four registers: $q(X)$ which holds the ancillas for X parity check measurements, $q(Z)$ which holds Z parity check ancillas, $q(L)$ for the left and $q(R)$ for the right data qubits.
We divide $n$ data qubits into the left and the right registers $q(L)$ and $q(R)$ of size $n/2$ each.  Each check operator acts on either two or three data qubits from $q(L)$ and two or three data qubits from $q(R)$.  In total, there are $2n$ qubits, with $n$ data qubits and $n$ ancillas to record  the syndrome of the parity checks.  
We label qubits in each register by integers $i=1,2,\ldots,n/2$.
We write $q(X,i)$ for the $i$-th qubit of the register $q(X)$ with similar notations for $q(L)$, $q(R)$, and $q(Z)$.
The matrices $A_p$ and $B_q$ from the MB code define the parity checks, and can be seen as  one-to-one map from the set $\{1,2,\ldots,n/2\}$ onto itself.

Note that within each round all operations act over non-overlapping sets of qubits.  Qubits from $q(Z)$ are always targets for CNOTS, such 
 that  $X$-type errors propagate from data qubits to check qubits in $q(Z)$. Simiarly,  $q(X)$ are always controls for CNOT for $Z$-type errors propagate from data qubits to check qubits in $q(X)$.

The syndrome measurement circuit is shown in Tab.~\ref{tab:syndromecircuit}, where we have the CNOT gate $\cnot{c}{t}$ with control qubit $c$ and target qubit $t$, 
$\initX{q}$ to intilize qubit $q$ in the state $\ket{+}$ state, 
$\initZ{q}$ to initialize qubit $q$ in the state $\ket{0}$ state, $\measX{q}$ to measure qubit $q$ in the $X$-basis and $\measZ{q}$ in the $Z$-basis, while 
$\idle{q}$ means the qubit $q$ is doing nothing.

\begin{table}[ht]
\begin{center}
\begin{tabular}{c|c||c|c}
\hline
Round  & Circuit & Round    & Circuit \\
\hline
\hline
1 & 
{\centering
\begin{minipage}{7cm}
	\begin{algorithmic}
	\State{}
	\For{$i=1$ to $n/2$}
	\State{$\initX{q(X,i)}$}
        \State{$\initZ{q(Z,i)}$}
        \State{$\idle{q(R,i)}$}
        \State{$\idle{q(L,i)}$}
	\EndFor
	\State{}
	\end{algorithmic}
\end{minipage}}
&  5 &
{\centering
\begin{minipage}{7cm}
	\begin{algorithmic}
	\State{}
	\For{$i=1$ to $n/2$}
	\State{$\cnot{q(X,i)}{q(R,B_1(i))}$}
	\State{$\cnot{q(L,B_2^T(i))}{q(Z,i)}$}
	\EndFor
	\State{}
	\end{algorithmic}
\end{minipage}}
 \\
\hline
2 & 
{\centering
\begin{minipage}{7cm}
	\begin{algorithmic}
	\State{}
	\For{$i=1$ to $n/2$}
	\State{$\cnot{q(X,i)}{q(L,A_1(i))}$}
	\State{$\cnot{q(R,A_3^T(i))}{q(Z,i)}$}
	\EndFor
	\State{}
	\end{algorithmic}
\end{minipage}}
& 6 & 
{\centering
\begin{minipage}{7cm}
	\begin{algorithmic}
	\State{}
	\For{$i=1$ to $n/2$}
	\State{$\cnot{q(X,i)}{q(L,A_3(i))}$}
	\State{$\cnot{q(R,A_1^T(i))}{q(Z,i)}$}
	\EndFor
	\State{}
	\end{algorithmic}
\end{minipage}}
\\
\hline
3 & 
{\centering
\begin{minipage}{7cm}
	\begin{algorithmic}
	\State{}
	\For{$i=1$ to $n/2$}
	\State{$\cnot{q(X,i)}{q(R,B_2(i))}$}
	\State{$\cnot{q(L,B_1^T(i))}{q(Z,i)}$}
	\EndFor
	\State{}
	\end{algorithmic}
\end{minipage}}
& 7 &
{\centering
\begin{minipage}{7cm}
	\begin{algorithmic}
	\State{}
	\For{$i=1$ to $n/2$}
	\State{$\measX{q(X,i)}$}
	\State{$\measZ{q(Z,i)}$}
	\State{$\idle{q(R,i)}$}
        \State{$\idle{q(L,i)}$}
	\EndFor
	\State{}
	\end{algorithmic}
\end{minipage}}
 \\
\hline
4 & 
{\centering
\begin{minipage}{7cm}
	\begin{algorithmic}
	\State{}
	\For{$i=1$ to $n/2$}
	\State{$\cnot{q(X,i)}{q(L,A_2(i))}$}
	\State{$\cnot{q(R,A_2^T(i))}{q(Z,i)}$}
	\EndFor
	\State{}
	\end{algorithmic}
\end{minipage}}
& 8 & 
{\centering
\begin{minipage}{7cm}

\end{minipage}}
\\
\hline
\end{tabular}
\end{center}
\caption{Depth-7 syndrome measurement circuit for weight 5 TB code $\dsl 30, 4, 5\dsr$.}
\label{tab:syndromecircuit}
\end{table}

To show that our syndrome circuit implements the correct checks, we use the stabilizer tableau formalism~\cite{aaronson2004improved}.   As $X$ and $Z$ operators are not mixed by CNOTs, we can consider the tableau for $X$ and $Z$ type Pauli separately.   

We now reproduce the formalism presented in Ref.~\cite{Bravyi_2024} for $X$ Pauli operators.  The corresponding tableau $T$ is a binary matrix of size $n \times 2n$ where each row of $T$ represents an $X$ stabilizer of the underlying quantum state. One can partition $T$ into four blocks for each qubit registers $q(X)$, $q(L)$, $q(R)$, and $q(Z)$.  One partitions $T$ into two blocks such that the top $n/2$ rows represent weight-$1$ check operators on $q(X)$ which are initially in the  $\ket{+}$ state, while the bottom $n/2$ rows are weight-$6$ check operators on data qubits. 
At the first round, all check qubits in $q(X)$ have been initialized in  $\ket{+}$, while the data qubits are in some logical state. The binary matrix is written as
\[
\begin{pmatrix}
I & 0 & 0 & 0 \\
0 & A & B & 0
\end{pmatrix}.
\]
Here $I\,{\equiv}\, I_{n/2}$ is the identity matrix. 
The circuit should give the transformation
\[
\begin{pmatrix}
I & 0 & 0 & 0 \\
0 & A & B & 0
\end{pmatrix}
\xrightarrow{\text{syndrome circuit}}
\begin{pmatrix}
I & A & B & 0 \\
0 & A & B & 0 
\end{pmatrix}
\]
such that the circuit maps a single-qubit $X$ stabilizer $X_j$ on a check qubit $j\,{\in}\, q(X)$ to a product of $X_j$ and the $j$-th $X$ check acting on data qubits is given by the $j$-th row of $H^X=[A|B]$.
The eigenvalue measurement of $X_j$ at the final round then reveals the syndrome of the $j$-th check operator.
The bottom $n/2$ rows should not change as the check operators of the code must be the same before and after the syndrome measurement.

One can verify that our circuit enacts this transformation, which can be done in a simple manner by removing unneeded notations~\cite{Bravyi_2024}
We denote each CNOT operation as $\cnotgate_M(a,b)$, where $a,b \,{\in}\, \{1, 2, 3, 4\} = \{q(X), q(L), q(R), q(Z)\}$, and $M \,{\in}\, \{A_1,A_2,A_3,B_1,B_2\}$. When $M^T$ is used in Tab.~\ref{tab:syndromecircuit}, one can instead write $M$ by renaming the variable as $i\,{\gets}\, M(i)$.

One can now write the syndrome checks compactly as
\begin{equation}
\begin{array}{cc}
\text{Round~1:} &  \cnotgate_{A_1}(1,2), \cnotgate_{A_3}(3,4) \\
\text{Round~2:} & \cnotgate_{B_2}(1,3), \cnotgate_{B_1}(2,4) \\
\text{Round~3:} & \cnotgate_{A_2}(1,2), \cnotgate_{A_2}(3,4) \\
\text{Round~4:} & \cnotgate_{B_1}(1,3), \cnotgate_{B_2}(2,4) \\
\text{Round~5:} & \cnotgate_{A_3}(1,2), \cnotgate_{A_1}(3,4) 
\end{array}
\label{syndromecheck}
\end{equation}

We now apply all rounds to verify that our circuit indeed implements the correct parity checks
\[
\text{Round~1:}
\begin{pmatrix}
I & 0 & 0 & 0 \\
0 & A & B & 0
\end{pmatrix}
\xrightarrow{\cnotgate_{A_1}(1,2)}
\begin{pmatrix}
I & A_1 & 0 & 0 \\
0 & A & B & A_1 B
\end{pmatrix}
\xrightarrow{\cnotgate_{A_3}(3,4)}
\begin{pmatrix}
I & A_1 & 0 & 0 \\
0 & A & B & A_3 B
\end{pmatrix}
\]

\[
\text{Round~2:}
\xrightarrow{\cnotgate_{B_2}(1,3)}
\begin{pmatrix}
I & A_1 & B_2 & 0 \\
0 & A & B & A_3 B
\end{pmatrix}
\xrightarrow{\cnotgate_{B_1}(2,4)}
\begin{pmatrix}
I & A_1 & B_2 & A_1B_1\\
0 & A & B & A_3B + AB_1 \\
\end{pmatrix}
\]

\[
\text{Round~3:}
\xrightarrow{\cnotgate_{A_2}(1,2)}
\begin{pmatrix}
I & A_1+A_2 & B_2 & A_1B_1 \\
0 & A & B & A_3B+AB_1 \\
\end{pmatrix}
\xrightarrow{\cnotgate_{A_2}(3,4)}
\begin{pmatrix}
I & A_1 +A_2 & B_2 & A_1B_1+A_2B_2 \\
0 & A & B & A_3B + AB_1+A_2B \\
\end{pmatrix}
\]

\[
\text{Round~4:}
\xrightarrow{\cnotgate_{B_1}(1,3)}
\begin{pmatrix}
I & A_1 +A_2 & B & A_1B_1+A_2B_2 \\
0 & A & B & A_3B + AB_1+A_2B \\
\end{pmatrix}
\]
\[
\xrightarrow{\cnotgate_{B_2}(2,4)}
\begin{pmatrix}
I & A_1 +A_2 & B & (A_1+A_2)B_2+A_1B_1+A_2B_2 \\
0 & A & B & A_3B+AB_1+A_2B+AB_2 \\
\end{pmatrix}
=
\begin{pmatrix}
I & A_1 +A_2 & B & A_1B_2+A_1B_1 \\
0 & A & B & A_1B \\
\end{pmatrix}
\]
Here, we used $B_1+B_2=B$, $2A_2B_2=0$, $A_2B+A_3B=A_1B+AB$ and $2AB=0$.
\[
\text{Round~5:}
\xrightarrow{\cnotgate_{A_3}(1,2)}
\begin{pmatrix}
I & A & B & A_1B_2+A_1B_1 \\
0 & A & B & A_1B \\
\end{pmatrix}
\xrightarrow{\cnotgate_{A_1}(3,4)}
\begin{pmatrix}
I & A & B & A_1B_2+A_1B_1+A_1B \\
0 & A & B & A_1B + A_1B \\
\end{pmatrix}
=
\begin{pmatrix}
I & A & B & 0 \\
0 & A & B & 0  \\
\end{pmatrix}
\]
where we used $A_1+A_2+A_3=A$ and $A_1B_1+A_1B_2=A_1B$. Thus, although we interleave $X$ and $Z$ checks, we end up with the correct parity check for $X$.
Similarly, one can verify that $Z$ checks are  also preserved, as well as the logical operators~\cite{Bravyi_2024}.

Note that Ref.~\cite{Bravyi_2024} proposed a syndrome measurement circuit for weight 6 codes in depth 8. This circuit can be adapted to weight 5 codes with depth 7 by a simple modification. In particular, from the original protocol~\cite{Bravyi_2024} ones removes round 5, else leave the circuit the same. However, we find that this syndrome measurement circuit for weight 5 does not preserve the code distance under circuit-level noise, i.e. $d_\text{circ}<d$. We also checked also re-arrangements of the circuit structure and found no improvement. We write the circuit out below in compact notation:
\begin{equation}
\begin{array}{cc}
\text{Round~1:} & \hfill \cnotgate_{A_1}(3,4) \\
\text{Round~2:} & \cnotgate_{A_2}(1,2), \cnotgate_{A_3}(3,4) \\
\text{Round~3:} & \cnotgate_{B_2}(1,3), \cnotgate_{B_1}(2,4) \\
\text{Round~4:} & \cnotgate_{B_1}(1,3), \cnotgate_{B_2}(2,4) \\
\text{Round~6:} & \cnotgate_{A_1}(1,2), \cnotgate_{A_2}(3,4) \\
\text{Round~7:} & \cnotgate_{A_3}(1,2) \hfill \\
\end{array}
\label{SCunitary_part}
\end{equation}

\section{Logical circuits}
A key challenge is to implement logical operations on the encoded logical qubits in a fault-tolerant way. 
Optimally, logical operations should be implementable transversally, i.e. each physical operation affects only one data qubit at a time. A slightly more relaxed condition is transversal operations combined with SWAPs, i.e. interchanging data qubit positions. This can be often implemented easily in many setups, e.g. ion traps or Rydberg atoms. 

Recently, a method based on automorphism groups to find such logical operations involving only transversal operations and SWAPs has been proposed~\cite{sayginel2024fault}. Using Ref.~\cite{sayginel2024fault} approach, we are able to find such logical operators for our codes.

\begin{figure}[htbp]
	\centering	
    \subfigimg[width=0.25\textwidth]{a}{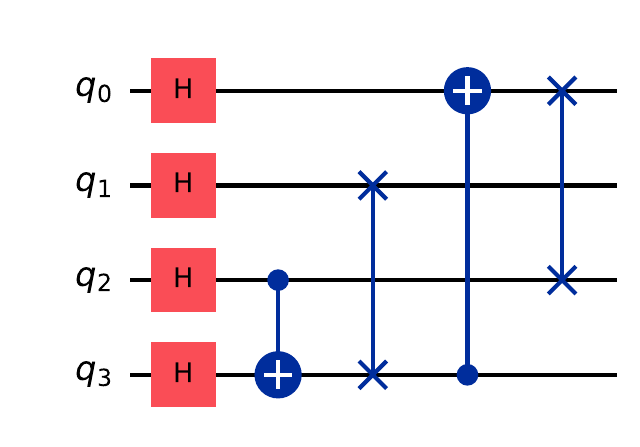}
    \subfigimg[width=0.28\textwidth]{b}{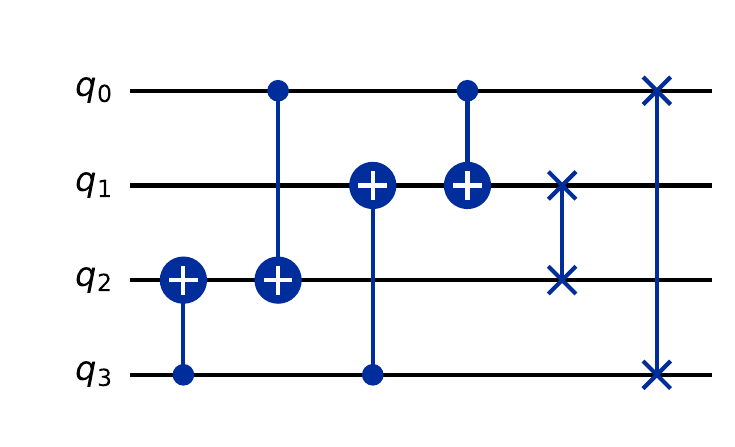}
    \subfigimg[width=0.28\textwidth]{c}{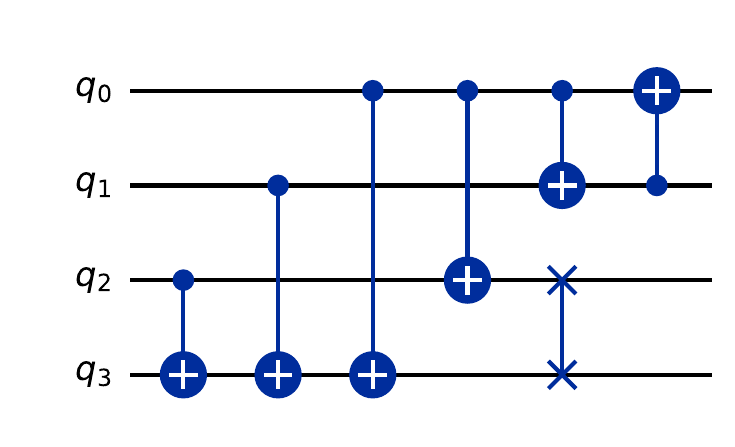}
    \subfigimg[width=0.14\textwidth]{d}{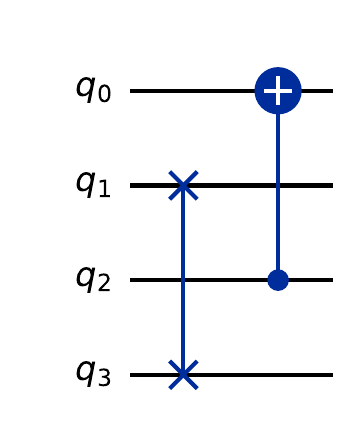}
    \subfigimg[width=0.28\textwidth]{e}{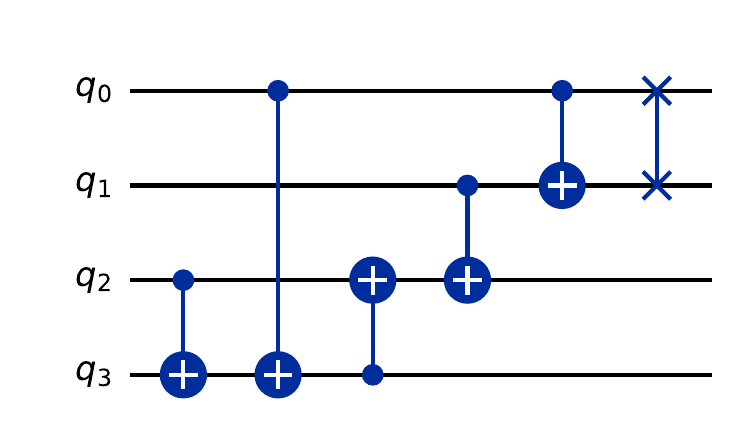}
    \subfigimg[width=0.35\textwidth]{f}{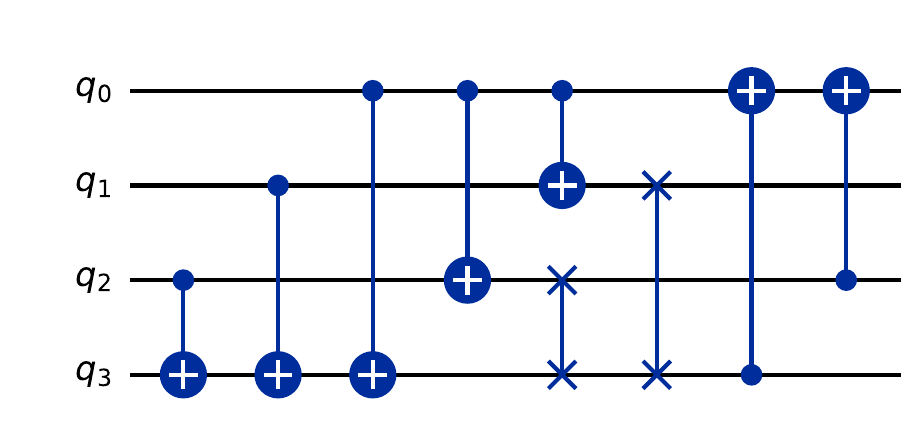}
    \subfigimg[width=0.25\textwidth]{g}{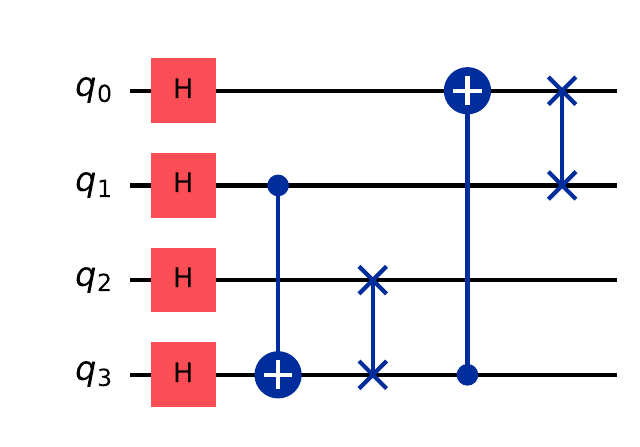}
	\caption{Seven different logical circuits implementable by transversal operations and SWAPs for the weight-5 $\dsl 30, 4, 5\dsr$ code.
	}
	\label{fig:logicals}
\end{figure}

For example, for our $\dsl 30, 4, 5\dsr$  weight 5 code we find 7 unique non-trivial logical operators. We plot them in Fig.~\ref{fig:logicals}.

\end{document}